\documentclass[useAMS,usenatbib]{mn2e}
\usepackage{graphicx}
\def\etal{{et al. }}

\def\kms{\ {\rm km\,s^{-1}}}

\title[Structure and dynamics of the galaxy cluster AC114]
{The structure and dynamics of the AC114 galaxy cluster revisited
\thanks{Based on observations made with ESO Telescopes at the La Silla Paranal Observatory (Chile) under
programme ID 083.A-0566.}}

\author[Dominique Proust \etal]
{Dominique Proust$^{1}$, Irina Yegorova$^{2}$, 
Ivo Saviane$^{2}$, Valentin D. Ivanov$^{2}$
\newauthor Fabio Bresolin$^{3}$ John J. Salzer$^{4}$ and Hugo V. Capelato$^{5,6}$\\
$^{1}$Observatoire de Paris-Meudon, GEPI, F92195 MEUDON, France\\
$^{2}$European Southern Observatory, Alonso de Cordova 3107, 
Vitacura, Casilla 19001, Santiago de Chile~19, Chile\\
$^{3}$Institute for Astronomy, 2680 Woodlawn Drive Honolulu, HI 96822 USA\\
$^{4}$Department of Astronomy, Indiana University, 727 East Third Street,
Bloomington, IN 47405, USA\\
$^{5}$Divis\~ao de Astrof\'isica, INPE-MCT, 12227-010 S\~ao Jos\'e dos Campos, S\~ao Paulo, SP, Brazil\\
$^{6}$N\'ucleo de Astrof\'isica Te\'oretica, Universidade Cruzeiro do Sul, rua Galv\~ao Bueno 868,
CEP 01506-200, S\~ao Paulo, SP, Brazil\\}

\begin{document}

\date{Accepted date / Received date; in original form}

\maketitle

\label{firstpage}

\begin{abstract}
We present a dynamical analysis of the galaxy cluster AC114 based on a catalogue of 524~velocities. Of 
these, 169 (32\%) are newly obtained at ESO (Chile) with the VLT and the VIMOS spectrograph. Data on
individual galaxies are presented and the accuracy of the measured velocities is discussed.
Dynamical properties of the cluster are derived. We obtain an improved mean redshift value
$z= 0.31665 \pm 0.0008$ and velocity dispersion $\sigma= 1893^{+73}_{-82}\kms$. A large velocity 
dispersion within the core radius and the shape of the infall pattern suggests that this part of the
cluster is in a radial phase of relaxation with a very elongated radial filament spanning $12000 \kms$. 
{ A radial foreground structure is detected within the central 0.5/h~Mpc radius, recognizable as a redshift group
at the same central redshift value}. We analyze the color distribution for this archetype Butcher-Oemler galaxy 
cluster and identify the separate red and blue galaxy sequences. The latter subset contains 44\% of confirmed members
of the cluster, reaching magnitudes as faint as $R_{f}$= 21.1 (1.0 magnitude fainter than previous studies).
We derive a mass $M_{200}= (4.3 \pm 0.7) \times 10^{15}$ M$_{\odot}$/h. In a subsequent paper we will utilize the spectral
data presented here to explore the mass-metallicity relation for this intermediate redshift cluster.
\end{abstract}


%\titlerunning{cluster of galaxies AC114}

\section{Introduction} \label{Introduction}

Redshift surveys of clusters of galaxies are needed to study their dynamical and evolutionary state.
In clusters, the mean redshift is a key ingredient in deriving distances, allowing the study of matter
distribution on very large scales. Analysis of the velocity { distribution} within clusters can lead to an
estimate of the virial mass, constraining models of the dark matter content. Dynamical mass estimates complement
measurements at other wavelengths, in particular those obtained through X-ray observations of clusters. However,
discrepancies between optical, spectroscopic and X-ray mass estimators are often found (e.g., Girardi \etal 1998, 
Allen 2000, Cypriano \etal 2005). Virial mass estimates rely on the assumption of dynamical equilibrium.
X-ray mass estimates also depend on the dynamical equilibrium hypothesis and on the still not well-constrained
intracluster gas temperature gradient (e.g., Leccardi \& Molendi 2008) although it is much better established
from both {\it Chandra} and {\it XMM} data analyses. Finally, mass estimates based on gravitational lensing are
considered more reliable than the others (e.g., Mellier 1999) because they are completely independent of the
dynamical status of the cluster. The drawback is that lensing cannot probe the mass profile beyond the virial
radius of clusters. The discrepancies among the methods may come from the non-equilibrium effects in the central
region of the clusters (Allen 1998).

In this paper, we build upon previous studies of the dynamical status of the cluster AC114 with the addition 
of a new set of velocities. The observations of radial velocities reported here are part of a program 
to study the mass-metallicity relation (MZR) in AC114 (Saviane \etal 2014, Saviane \etal 2015
{\it in preparation}), which is a powerful diagnostic of galaxy evolution, as first shown by Larson (1974).
Note that the sensitivity of available instruments at 10m-class telescopes does not allow for the accurate
determination of the MZR at the highest redshifts, but for AC114 at $z\sim0.32$ we are able to measure abundances
reliably.

The AC114 cluster (R.A.=22h58mn52.3s Dec=$-34^{o} 46' 55''$ J2000) is classified as Bautz-Morgan type II-III by
Abell \etal (1989). It is the archetype object of the Butcher-Oemler effect with a higher fraction of blue,
late-type galaxies than in lower redshift clusters, rising to $60\%$ outside the core region
(Couch \etal 1998, Sereno \etal 2010).

AC114 has 724~galaxies listed in {\it Simbad} in an area of $\sim 10' \times 10'$,
and 585 of them are classified as emission-line galaxies. It has been observed several times. For example,
Couch \& Sharples (1987) derived 51~velocities (42~cluster members), Couch \etal (2001) with 51 $H_{\alpha}$
emitting cluster members, and more recently in the 2dF survey (Colless \etal 2003) and the 6dF survey 
(Jones \etal 2009). A total of 308~velocities are available in the NED database within a 10~arcmin radius,
348~within 20~arcmin and 414 within 30~arcmin. AC114 has a compact core dominated by a cD galaxy and a
strong lensing power with several bright arcs and multiple image sources (Smail \etal 1995,
Natarajan \etal 1998, Campusano \etal 2001). It is a hot X-ray emitter (kT= 8.0~keV) with an irregular
morphology. Below 0.5~keV, the X-ray emission is dominated by two main components: the cluster, roughly
centered on the optical position of AC 114, and a diffuse tail, extending almost 400~kpc from the cluster center to the
southeast. The cD galaxy is shifted with respect to the centroid of the X-ray emission but aligned in the general direction
of the X-ray brightness elongation (see de Filippis \etal 2004). An extensive study of AC114 based on a multiwavelength
strong lensing analysis of baryons and dark matter from Sereno \etal (2010) provided evidence of dynamical
activity, with the dark matter distribution being shifted and rotated with respect to the gas, following the galaxy
density in terms of both shape and orientation. Based on the lensing and X-ray data, they argue that the cluster
extends in the plane of the sky and is not affected by the lensing over-concentration bias. As our observations
increase by 32\% the data available for this cluster, we expand the dynamical analysis of AC114 in the present
paper. The mass-metallicity relation is developed in Saviane \etal (2014) and Saviane \etal (2015 {\it in preparation}).

We present in Sec.~2 the details of the observations and data reduction. In Sec.~3 we discuss the distribution and
the velocity analysis of the cluster galaxies. Sec.~4 describes the kinematical structures of AC114 and Sec.~5 makes
the analysis of the distribution in color and the Butcher-Oemler effect while in Sec.~6 we analyse
the dynamical mass determinations of AC114.  We summarize our conclusions in Sec.~7. We adopt here,
whenever necessary, H$_{0}$= 100$h~\kms$Mpc$^{-1}$, $\Omega_{M}$ = 0.3 and $\Omega_{\Lambda}$ = 0.7.

\section{Observations and Data Reductions}\label{Observations and Data Reductions}

The observations were carried out in service mode with the VIsible MultiObject Spectrograph
(VIMOS; Le F\`evre \etal 2003) mounted on the Very Large Telescope, ESO (083.A-0566). 
VIMOS is a visible light (360 to 1000 nm) wide field imager and multiobject spectrograph mounted 
on the Nasmyth focus~B of UT3 Melipal. The instrument is made of four identical arms each with 
a field of view of $7' \times 8'$, and a 0.205" pixel size. There is a $\sim2'$ gap between 
the quadrants. Each arm is equipped with 6~grisms providing a spectral resolution range
from $\sim 200-2500$ and with one EEV 4k $\times$ 2k CCD. Two dispersion modes were used: 
MR grating (spectral resolution of 580 for a 1'' slit, over 500 - 1000~nm spectral range) 
and HR-red grating (spectral resolution of 2500 for a 1'' slit, over 630 - 870~nm spectral range). 
The galaxy selection was made from the pre-imaging frames of the cluster. To construct the masks, initially, 
the SIMBAD catalogue was used without imposing any restriction criteria. As a first step we identified 
already known galaxies of the cluster. Then we selected the non-stellar objects in this region by eye in order 
to punch a maximum number of slits in each of the 4~quadrants. Such a visual inspection allows 
to discriminate between extended objects and stars. 

\begin{figure}
\begin{centering}
\includegraphics[angle=-90,width=1.0\columnwidth]{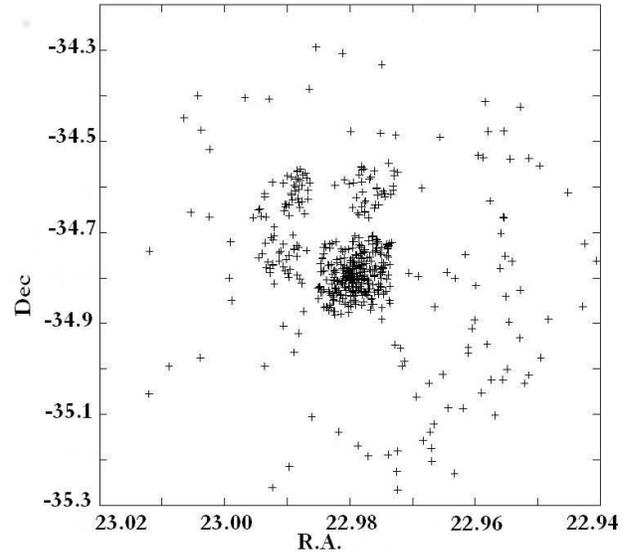}
\par\end{centering}
\caption{Map of the AC114 galaxies with observed velocities. The 4~central clumps are VIMOS observations
at the VLT while galaxies spread in the field are from the literature.}
\end{figure}

The 14~awarded hours of observations were done on 2009 August, 16 and 21 and September 17, 21 and 25
(see Table~1). Execution time for each observing block was 60~minutes, including 22.5~minutes of
overheads. Observing blocks were prepared for both MR and HR-red grisms. For each grism observing block
was repeated seven times. The total integration time on target galaxies is 4.37 hours for each grism. 

\begin{table*}[]
\caption{Observing blocks of AC114.} %\label{tbl1}
\small
\begin{tabular}{cccccl}
\hline
$OB_{ID}$  &  Date & OB start$^{\rm a}$&Exp. time$^{\rm b}$  &Filter  & Grism \\
\hline
382919 & 2009-09-17 &02:37& 2250&GG475 & HR red \\
382920 & 2009-09-17 &03:32& 2250&GG475 & HR red \\
382921 & 2009-09-17 &04:11& 2250&GG475 & HR red \\
382922 & 2009-09-17 &04:50& 2250&GG475 & HR red \\
382923 & 2009-09-17 &05:28& 2250&GG475 & HR red \\
382924 & 2009-09-17 &06:07& 2250&GG475 & HR red \\
382925 & 2009-09-21 &03:54& 2250&GG475 & HR red \\
382926 & 2009-08-16 &05:25& 2250&GG475 & MR \\
382927 & 2009-08-16 &06:15& 2250&GG475 & MR \\
382928 & 2009-08-16 &06:55& 2250&GG475 & MR \\
382929 & 2009-08-21 &07:24& 2250&GG475 & MR \\
382930 & 2009-08-21 &06:11& 2250&GG475 & MR \\
382931 & 2009-09-25 &04:02& 2250&GG475 & MR \\
382932 & 2009-09-25 &04:53& 2250&GG475 & MR \\
\hline
\end{tabular}

$^{\rm a}$ {\scriptsize  beginning of the observation block, UT time}

$^{\rm b}$ {\scriptsize exposure time of the science part, in seconds}
\end{table*}

Figure~1 shows the projected map in R.A. and Dec of our program galaxies in AC114 out to a velocity
$cz= 120000\kms$. Note that most of the objects in the central part represent 4~clumps corresponding to the
4~VIMOS quadrants of a VLT mask. The other galaxies spread in the field are objects already observed with velocities
published in the literature.

The data reduction was carried out independently at ESO-Santiago and Paris-Meudon observatory in order to obtain wavelength
and flux calibrated spectra. Thanks to the double reduction of the data, we could compare the quality of the results.
At ESO-Santiago we used the VIMOS pipeline to reduce the data. We reduced separately the data taken with the grism HR
red and MR. Each scientific frame was bias subtracted and flat field corrected, the cosmic rays were removed, and sky
emission lines subtracted (the same procedure was done for the standard stars).
The spectra were wavelength calibrated and seven frames for each VIMOS quadrant were combined.
This gave us S/N = 20 per pixel on the average at the continuum level on the final spectra. Since the continuum of most of the
galaxies is too weak we did not use the spectra extracted by the pipeline. Instead, we extracted one-dimensional
spectra with MIDAS.

At Meudon, we independently reduced the data with the MULTIRED package (Le F\`evre \etal 1995) of IRAF\footnote{IRAF
is distributed by the National Optical Astronomy Observatories, which are operated by the Association of Universities
for Research in Astronomy, Inc., under cooperative agreement with the National Science Foundation.} performing the
following steps in sequence for each slit:

- Extract small 2D postage-stamp images corresponding to one slit from the two dimensional spectra of the object and the
corresponding wavelength calibration and flat field from the full 4k $\times$ 2k pixel images.

- For each two-dimensional spectrum correct for flat field (pixel to pixel variation) and subtract sky
emission: the sky is fitted with adjustable low-order polynomials, and subtracted along the slit for
each wavelength element. A treatment of the zero-order position was also added: areas on the two-dimensional
spectra with a zero order could be corrected independently from the rest of the spectrum, if needed.

- Combine all the corrected two-dimensional spectra of a given object with average or median scheme using
sigma-clipping rejection. This removes most of the cosmic-ray events, although in some circumstances, the brightest
events can still partially remain.

- Extract a one-dimensional spectrum of the arc-lamps and cross-correlate with a reference arc-lamp
spectrum to produce an initial wavelength solution. The fit was then adjusted, if necessary. This produces a
unique pixel/wavelength transformation for each slit.

- Extract a one-dimensional spectrum from the corrected two-dimensional spectrum for each object of interest
in the slit by averaging along the wavelength axis.

- Wavelength and flux calibrate the one-dimensional object spectrum. We observed the F-type standard star
LTT1788 (V= 13.16, B-V= +0.47) and the DA-type LTT7987 (V= 12.23, B-V= +0.05) from Hamuy \etal (1992, 1994).  

- Plot the corrected and calibrated one-dimensional spectrum and display all the corrected two-dimensional
spectra together with the averaged two-dimensional spectrum.  Line identification for redshift
measurement can then proceed.

{ Because of the lack of CCD photometry for this cluster, we collected UK-J $B_{j}$ and ESO-R or POSS-I E $R_{f}$
magnitudes from superCOSMOS (Maddox \etal 1990a, 1990b); The photometric accuracy of these data is typically 
0.3~mag with respect to external data for $m \succ$~15. Colours (B-R or R-I) are externally accurate to 0.07~mag at 
$B_{j}$= 16.5 rising to 0.16 mag at  $B_{j}$ = 20. $B_{j}$ and $R_{f}$ magnitudes are listed in Table~2.} 
Looking at this table, the faintest objects for which superCOSMOS provides an apparent magnitude estimate have
magnitudes $B_{j}$= 22.6 and $R_{f}$= 21.1. However it is also clear that luminosities are not available
for roughly fifty percent of our targets. To overcome this limitation, we obtained a relation between the
integrated flux of our spectra and the $R_{f}$ magnitude, and then the relation was applied to get a first estimate
of the $R_{f}$ magnitude for objects that do not have a luminosity estimate.

The flux integral was computed simply as $\Sigma f\times\delta\lambda$, where $f$ is the flux in units of
10$^{-16}$~erg~cm$^{-2}$sec$^{-1}$\AA$^{-1}$ and $\delta\lambda=2.6$\AA\ is the spectral bin. The flux integral
was limited to the common spectral range of the spectra, which is from 4897.5\AA\ to 9785.5\AA. The relation 
between the integrated flux and $R_{f}$ is shown in Fig.~2. Assuming that the relation is linear, with a
preliminary fit we found that the slope of the regression is $\sim 2.7$, which suggests that our integrated
flux is close to what is measured by the R-band photometry. Therefore we imposed the relation to be
$R_{f}=-2.5\,\log f+R_{0}$ and we found $R_{0}= 25.62$, with a dispersion of 0.51~dex around the fit.

The dispersion is quite satisfactory for an order-of-magnitude estimate of galaxian luminosities, and
it is likely due to object-to-object variations in slit losses, night-to-night differences in the flux
calibration of the VIMOS spectra, and variable sky transparency in the course of the observations.

Using the relation above, for each galaxy an $R_{comp}$ magnitude can be computed, and luminosity distributions
can be produced. In Fig.~3 we show such distributions for 25~galaxies that already have a spectrum in the literature
as flagged by an entry in the NED database\footnote{The NASA/IPAC Extragalactic Database (NED) is operated by the Jet
Propulsion Laboratory, California Institute of Technology, under contract with the National Aeronautics and Space
Administration.}, and for 138~galaxies added with the present work. { No selection was done;
this work covers a larger area than older datasets}. The apparent magnitude distribution of galaxies with existing
spectra turns over at $R_{comp} \sim 19$ while that of our new sample keeps growing down to $R_{comp} \sim 20$.
Below $R_{comp} \sim 19$ the literature sample thus suffers from incompleteness compared to our new sample, which
reaches $\sim 1$ magnitude deeper than existing data. { The mismatch between the $R_{f}$ and
$R_{comp}$ sensitivity/integration bands may also contribute to the dispersion in the relation}.    

\begin{figure}
\begin{centering}
\includegraphics[width=0.5\textwidth]{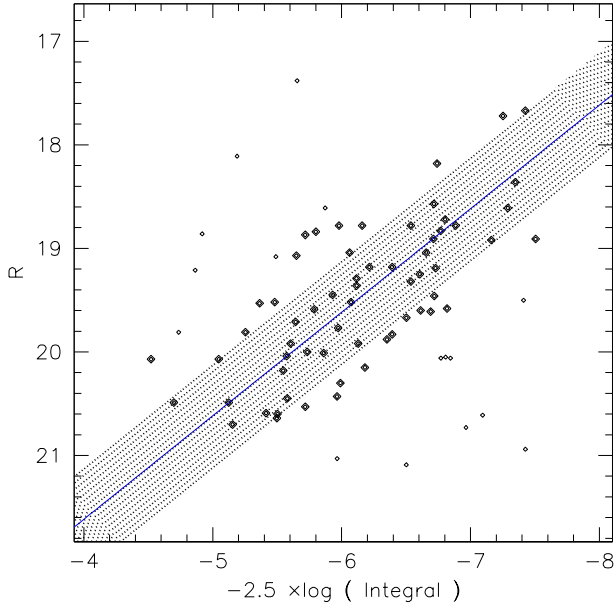}
\par\end{centering}
\protect\caption{Relation between $R_{f}$ magnitudes and the flux integral described
in the text. Objects identified by larger symbols were considered in the linear
regression, which is shown by the straight line. The shaded area represents
the $\pm1\sigma$ dispersion around the fit. \label{fig:Relation-between-}}
\end{figure}

\begin{figure}
\begin{centering}
\includegraphics[width=0.5\textwidth]{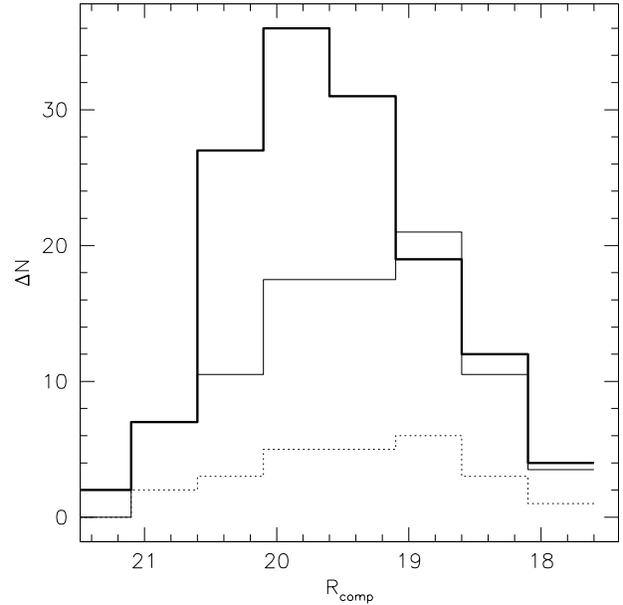}
\par\end{centering}
\protect\caption{Luminosity distribution of galaxies with existing spectra
(dotted-line histogram) and of galaxies added by the present study (thick-line histogram).
To better compare the two samples, the thin-line histogram has been scaled from the dotted-line
one so that the sum of counts in the three brightest bins is the same as that of the thick-line
histogram. { This scaling factor of 3.5 almost coincides with the ratio of the area surveyed by us
and the area defined by literature data}. R-band magnitudes were computed as described in the text.
\label{fig:The--luminosity}}
\end{figure}

In order to study the completeness of our catalogue, we compared it to the corresponding superCOSMOS plate\footnote
{This research has made use of data obtained from the superCOSMOS Science Archive, prepared and hosted by the Wide
Field Astronomy Unit, Institute for Astronomy, University of Edinburgh, which is funded by the UK Science and
Technology Facilities Council.}(Hambly \etal 2001) as our faintest program galaxies are close to the superCOSMOS 
limiting magnitude ($\simeq B_{j}$=23.0 and $R_{f}$=21.5). In the core radius $R_{c}$= 1.02~arcmin
(0.27/h~Mpc see Couch \etal 1987), all of the 8~galaxies have velocities. In the Abell radius computed above
$R_{Abell}$= 5.431~arcmin (1.44/h~Mpc), 109 out of 173 galaxies (63\%) have velocities and in the virial radius
$R_{v}$= 3.0/h~Mpc (see Couch \etal 2001) 340 out of 724 galaxies (47\%) have velocities.

Radial velocities have been determined using the cross-correlation technique (Tonry 1979) implemented in XCSAO
task of the RVSAO package (Kurtz 1991, Mink 1995) with spectra of radial velocity standards of late-type stars 
(Pickles 1998) and previously well-studied galaxies (Pickles 1985). The values of their R statistics 
(defined as the ratio of the correlation peak height to the amplitude of the antisymmetric noise) are listed
in Table~2 along with the measured velocities and there formal uncertainties.  For spectra with $R~\prec~3.0$
the measured velocity was considered unreliable and was not used, except for emission-line objects where the 
velocity was obtained using the EMSAO task implemented in the RVSAO package.

A total of 176 of the 200~observed spectra had S/N high enough to measure a useable velocity. Of these,
169~were galaxies, while 7~of them were found to be stars. Note that for 4~galaxies we derived two independent
redshifts: object~45 in quadrant~1 with redshifts respectively 0.34461 and 0.35159, object~1 in quadrant~2
(0.25782 and 0.33069), and in quadrant~4 object~9 (0.25733 and 0.29095) and object~55 (0.17199 and 0.31776);
after a visual inspection, these 4~targets seem to be composed by two objects in the line of sight.
With our observations, we increase the number of galaxies with velocities in the field of AC114 by 32\%.  
Most of these are situated in the central region of the cluster. The new redshift values with their individual error 
measurements are published in Table~2 and correspond to the highest R-value obtained from the cross-correlations.
We have constructed a velocity catalogue for AC114 with a total of 524~galaxies, where 169 are new measured 
velocities and 355~galaxies are from NED.

\begin{figure*}
\begin{centering}
\includegraphics[angle=-90,width=1.6\columnwidth]{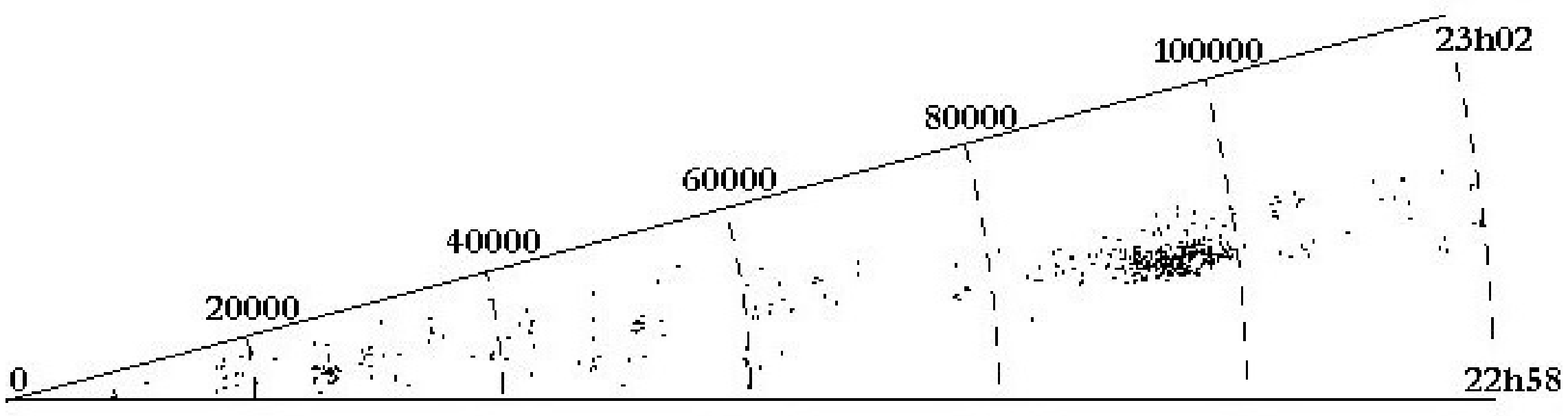}
\includegraphics[angle=-90,width=1.6\columnwidth]{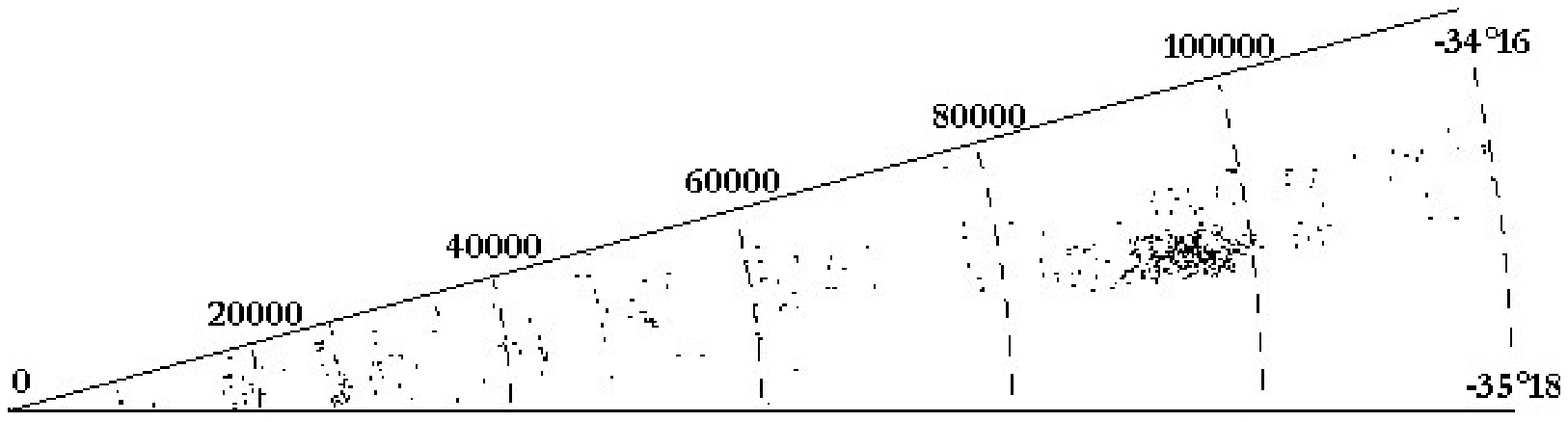}
\par\end{centering}
\caption{Wedge diagrams in R.A. and Dec for both the newly observed and literature 
galaxies until $120000\kms$.}
\end{figure*}

The contents of Table~2 are as follows:
\begin{enumerate} 

\item number of the object in each quadrant from slit position;

\item right ascension (J2000);

\item declination (J2000);

\item UK-J $B_{j}$ magnitude from superCOSMOS (Maddox \etal 1990a,b);

\item ESO-R or POSS-I E $R_{f}$ magnitude from superCOSMOS (Maddox \etal 1990a,b);

\item $R_{comp}$ computed magnitude from spectra;

\item redshift;

\item redshift error;

\item R value from the cross-correlation (Tonry 1979);

\item notes.
\end{enumerate}

Figure~4 shows the wedge diagrams in R.A. and Dec for both set of observed and NED galaxies out to
$120000\kms$ with 468~galaxies; in the $\sim10\arcmin\times10\arcmin$ central area of the cluster we 
have 189~velocities. Among the 169 velocities of table~2, 26~galaxies have previously published velocities 
listed in NED. Figure~5 shows the velocity comparison for these 26~galaxies. It can be seen that NED velocities
tend to be larger than those measured in this work, although this is just a $0.7\sigma$ effect. The dispersion
is larger than our measurement errors, which is likely due to the non-homogeneity of the NED sample.
We note a large velocity difference for the galaxy~56 in quadrant~4 of our table~2 (98377$\kms$ and 97642$\kms$
from Couch \etal 1998). As our correlation coefficient is weak (R=3.31) and our velocity error large (132$\kms$)
we give preference to their value.

\begin{figure}
%\begin{centering}
\includegraphics[width=1.0\columnwidth]{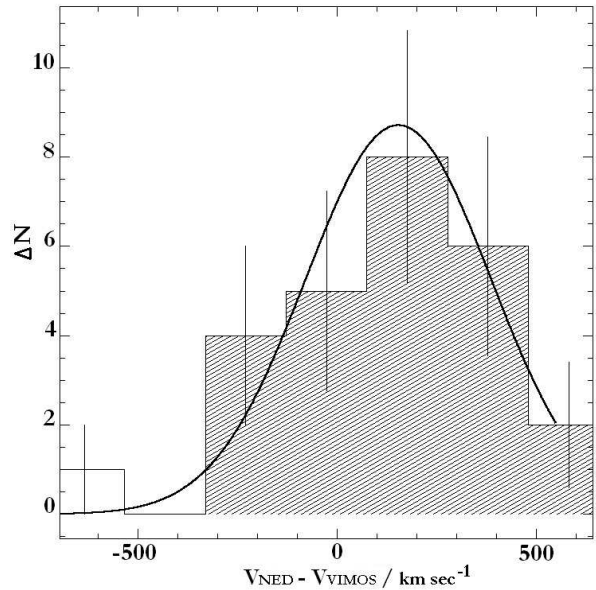}
%\par\end{centering}
\caption{Velocity comparison between 26~velocities in common with the present work (VIMOS) and the NED database.
After discarding the outlying galaxy with a very low NED velociy, the $v_{{\rm NED}}-v_{{\rm VIMOS}}$
distribution (shaded histogram) resembles a Gaussian one, albeit with a slight excess on the small values side.
Velocity differences are distributed around the average 154$\kms$ with a dispersion of 232$\kms$. The normal
distribution with these parameters and the same area of the histogram is shown by the solid curve.}
\end{figure}

Figure~6 shows the sky distribution of emission-line galaxies (blue dots) and of passive galaxies (red dots)
in our newly measured spectra for six different redshift ranges. There are 21~emission-line galaxies
at the redshift of AC114 out of a total of 86~cluster members. The bottom panel shows that the fraction of
emission-line galaxies has a dip at the redshift of AC114, and that these galaxies tend to avoid
the cluster center. Error bars in that panel show the 95\% confidence limits, calculated according to 
formula 3.27 in Feigelson \& Babu (2013).\footnote{We have assumed that for our relatively large samples,
the binomial distribution can be approximated by a normal distribution, so that $z_{\alpha/2}=$~1.96 in the formula.
No confidence intervals were computed for the lowest and the highest redshift bin where the Gaussian 
approximation to the binomial distribution does not hold.}
At such a redshift, all important emission lines still fall in the optical range, and the universe is $\simeq$
70\% its current age, so we can expect a factor 1.4 increase in $Z$ since that time, or 0.14~dex in $[m/H]$
(Gullieuszik \etal 2009, Leaman \etal 2013). A discussion concerning the mass-metallicity relation from this
new set of emission-line galaxies is developed in Saviane \etal (2014) and Saviane \etal (2015 {\it in preparation}).

\begin{figure*}
\begin{centering}
\includegraphics[width=1.2\columnwidth]{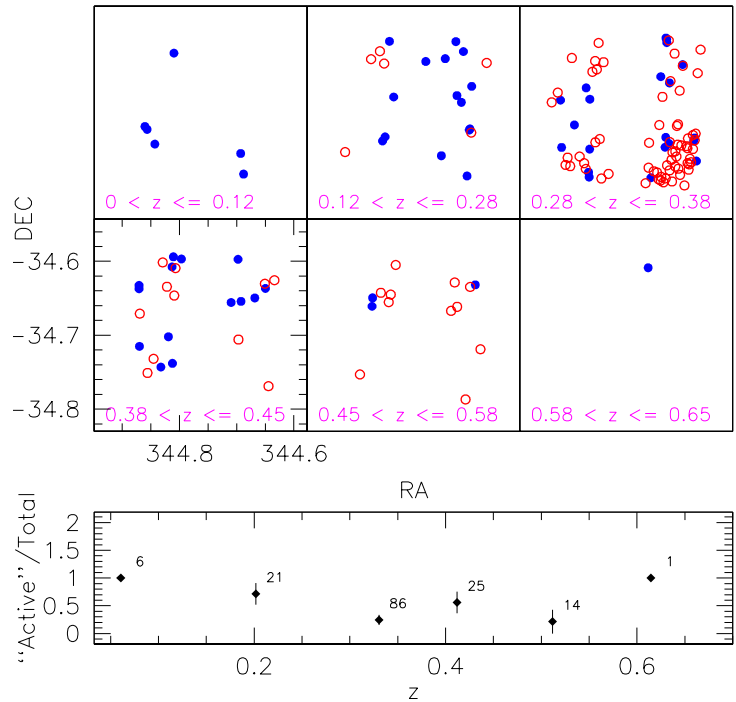}
\par\end{centering}
\caption{The top panels show the sky distribution of emission-line galaxies (blue dots) and of passive
galaxies (red dots) for six redshift groups. The bottom plot shows the ratio of emission-line-to-total number of
galaxies as a function of redshift. Error bars represent the 95\% confidence limits, and numbers near each point
give the total number of galaxies for each redshift interval.}
\end{figure*}

\section{Velocity analysis}\label{Velocity analysis}

Figure~7 shows the redshift distribution of the spectroscopic sample within an area of $\simeq$~1.0
square degree centered on AC114. The large central peak at z $\sim$ 0.315 corresponds to the main cluster. Other
structures are seen, most for $z \leq$ 0.25, as well also for $z \sim$ 0.41. Note the important contamination
by fore - and background objects along the line of sight: 45\% of the galaxies in our redshift catalogue are
non-members of AC114. Couch \& Sharples (1987) found it to be 18\% from 51~galaxies observed in a field of
only $5\times5$~arcmin. In the same field we have 114~redshifts with a contamination rate of 23\% which is not
significantly different.

We used the ROSTAT routines (Beers \etal 1990) to analyze the galaxy velocity distribution of AC114. The
dominant kinematical structure shown on Figure~7 contains 265~galaxies between $z$=0.295 and $z$=0.34,
including 77~new velocities. The rest-frame velocities $v= c(z-\overline{z})/(1+\overline{z})$ range from
$-4312\kms$ to $3955\kms$ at $\overline{z}= 0.31665^{+0.00072}_{-0.00082}$. The large rest-frame dispersion
comes out as ${\sigma}= 1893^{+73}_{-82}\kms$ suggesting that the cluster could be in a radial phase of
relaxation with the presence of fore and background structures. The very elongated radial filament spanning 
$12000 \kms$ seen in Figure~4 is a manifestation of the familiar {\it finger-of-god} effect due to this 
large internal dispersion velocity. Note that all the normality tests included in the ROSTAT package fail to
reject the null hypothesis of a Gaussian distribution for this sample.

We can deduce that AC114 has an Abell radius defined by $R_{Abell}(arcmin) \sim 1.72/\overline{z}$,
which gives $R_{Abell}$= 5.431~arcmin (1.44/h~Mpc) with an angular size distance of 909/h~Mpc and a scale of
264.54 Kpc arcmin$^{-1}$ (and a luminosity distance of 1560/h~Mpc from the mean redshift corrected to the
Reference Frame defined by the 3K CMB). Note that Couch \& Sharples (1987) found a dispersion 
${\sigma}= 1649^{+217}_{-156}\kms$ from a set of 42~galaxies in a square region of $5\times5$~arcmin
($1.32 \times 1.32$/h~Mpc) and Mahdavi \& Geller (2001) with ${\sigma}= 1660^{+128}_{-106}\kms$ within a radius
of 4.45~arcmin (1.18/$h_{50}$~Mpc). Finn \etal (2004) give a dispersion of ${\sigma}= 1390\kms$ with
$r_{200}$= 1.87/$h_{100}$ (9.54~arcmin) and Just \etal (2010) obtain similar values with $r_{200}$= 3.98/$h_{70}$~Mpc
radius with 196~galaxies: $\overline{z}= 0.31500$ and ${\sigma}= 1889^{+81}_{-74}\kms$. Figure~8 shows the rest-frame
velocity histogram between $89000$ and $101000\kms$ and the 
gaussian with $\sigma= 1893\kms$.

\begin{figure*}
\begin{centering}
\includegraphics[width=1.3\columnwidth]{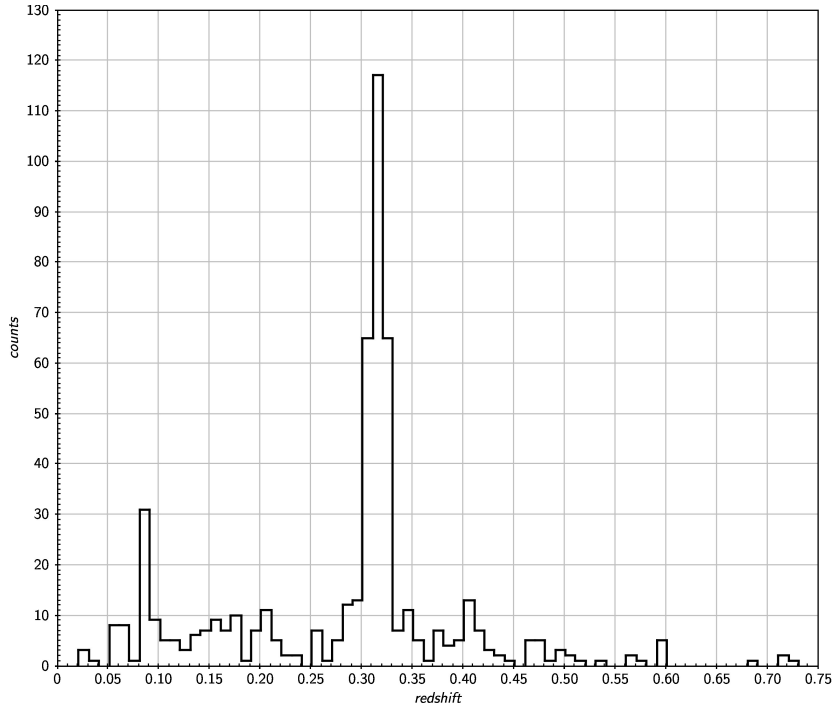}
\par\end{centering}
\caption[]{The redshift distribution within an area of $\sim$ 1.0 square degrees centered on the AC114 cluster.}
\end{figure*}

The brightest cluster member (BCG; R.A.=22h58mn48.4s Dec=$-34^{o}$48'08'' J2000) has a { redshift 
$z= 0.31691 \pm 0.00029$}, very close to the average redshift of the cluster. This galaxy is not
infrared-luminous compared with other massive early-type galaxies, suggesting that the cluster environment has
little influence on the infrared luminosities of these galaxy as discussed by Martini \etal (2007).

\section{Kinematical structures}\label{Kinematical structures}

\subsection{Method}
In order to get more precise identifications of these  structures, we applied the  gap technique throughly 
discussed by Katgert \etal (1996). These consists in fixing a maximum allowed gap size in the rest-frame peculiar velocity
space as a condition for galaxies to belong to a given structure. Adami \etal (1998) pointed out that gap sizes should not 
be kept constant, but should depend on the size $N_{S}$ of redshift sample since it is more likely to find large gaps in
sparse data-sets than in the richer ones, whereas a fixed gap would tend to overestimate the number of groups when
$N_{S}$ is small. Based on simulations of Gaussian distributions with varying number of objects they define the 
{\it density gap} such that:

\begin{equation}  
{\Delta v_{max} = \delta_{gap} \cdot f(}N_{S})
\end{equation}

\noindent with

\begin{equation}
{f(}N_{S}\mathrm{) = (1 + exp(-(}N_{S} \mathrm{-6)/33)} 
\end{equation}

\noindent and $\delta_{gap} = 500\kms$. In redshift space this translates to: 

\begin{equation}
{\Delta z_{max} = (500/}c\mathrm{)(1+z_{cl})f(}N_{S})
\end{equation}

\noindent where $\mathrm{z_{cl}}$ is the mean redshift of the candidate structure and $c$ is the speed of light in $\kms$.

We ran this gapping procedure for each of the candidate structures observed in the histogram displayed in Figure~7.
Following Lopes \etal (2009; see also Ribeiro \etal 2013) we start by first looking at their densest central region,
assumed to be circular with 0.5/h~Mpc radius. We assume the central position of the main cluster as the initial
guess for the calculations and further refine the central position of each candidate structure after collecting the
galaxies belonging to the $z$-groups associated to it and re-starting the whole procedure. The $z$-groups detected 
by the gapping have their mean (bi-weighted) redshift $\bar{z}_{\nu}$ estimated and ranked accordingly to their distance
to the nominal redshift of the candidate kinematical structure, $\delta_{\bar{z}} \equiv |\bar{z}_{\nu} - \mathrm{z_{cl}}|$,
with the most nearby being identified to it, providing:
% whenever satisfying a maximum distance:

\begin{equation}
\delta_{\bar{z}} \leq \mathrm{\Delta v_{max} (1+z_{cl})/}c
\end{equation}

\noindent with $\Delta v_{max} = 1000\kms$, allowing for the uncertainties and biases affecting the estimate of $\mathrm{z_{cl}}$.
Normally only one group survives this criteria and we then assume $\mathrm{z_{cl}}= \bar{z}_{\nu}$ for this candidate structure.
{ A new central sky position is calculated from the positions of the galaxies belonging to this $z$-group and the whole
procedure is iterated once. Next we search for the extreme values $z_{min}$, $z_{max}$ of the redshift distribution of the
structure and then select all galaxies within 2.5/h~Mpc from its center according to:

\begin{equation}  
|z_{i} - \mathrm{z_{cl}}| \leq max \{ (\mathrm{z_{cl}}-z_{min}),(z_{max}-\mathrm{z_{cl}}) \}
\end{equation}

We gauge the significance of the kinematical structures identified so far by bootstrap re-sampling the full redshift distribution
found inside the 0.5/h~Mpc radius around its center and re-applying the whole procedure exactly as before. We do that $N_{boot}=1000$
times and then check the probability of detection of the group:

\begin{equation}
p (\bar{z}_{\nu}) = \frac{N_{+}(\bar{z}_{\nu})}{N_{boot}}
\end{equation}

\noindent where  $N_{+}(\bar{z}_{\nu})$ is the number of positive detections for the group $\bar{z}_{\nu}$.}

Finally, in order to reject possible interloper galaxies we apply the "shifting gapper" technique 
(Fadda \etal 1996, Lopes \etal 2009), which consists on the same gapping procedure as described above but applied
within radial bins of width $\geq 0.4/h$~Mpc containing at least 15~galaxies. The maximum redshift gap is now fixed at:

\begin{equation}
{\Delta z_{max} = (300/}c{)(1+\mathrm{z_{cl}})f(}N_{S}) 
\end{equation}

and we require that the mean redshift of the group does not deviate from $\mathrm{z_{cl}}$ by more than 
$\delta_{\bar{z}} \leq 300(1+\mathrm{z_{cl}})/c$. If no $z$-group is found satisfying these requirements the procedure stops
and the aperture radius of the structure is identified to the internal radius of the bin. 

\subsection{Results}

We limited our analyses to the interval 0.25 $\leq z \leq$ 0.5, which contain the main peak due to AC114 as well
as neighbors { foreground and background} structures which may or may not interact with the main cluster.
Table~3 displays the main parameters of the kinematical structures we have found. Besides the main cluster AC114,
which dominates Figure~7, we have been able to detect two others secondary structures although at low levels of significance. 
{ Figure~9 (upper panel) shows the resulting redshift distribution of these structures and Figure~9 (lower panel) shows
with different symbols the projected positions of their galaxies}.

{ The foreground structure AC114 01 seems to be much more relevant for our discussion on the evolutionary status of the
main cluster. Although being detected as a redshift group of only 4 galaxies within the central 0.5/h~Mpc radius, and as so
suggesting to be a mere statistical fluctuation of the phase space distribution of galaxies, we argue that, besides the 77\% 
of chances of detection (on 1000 bootstrap trials), this structure is still recognizable as a redshift group at the same central
redshift value (within $c\delta_{z} \sim 300\kms$) at projected distances far from its very center. In fact the shifting gapper
was able to follow it out to $3.03/h_{73}$~Mpc which is $\sim$90\% of the limiting radius 2.5/h~Mpc. Given their proximity in
redshift space ~-~ almost certainly reflecting their relative space configuration ~-~ it is very difficult to completely 
disentangle the redshift distribution of AC114 01 from that of the main cluster. Nevertheless we have been able to make
roughly estimates of the population of interloper galaxies of AC114 01. In principle these should be constituted else by 
galaxies in their way of being torn from AC114 01 by the main cluster or by galaxies already dynamically linked to it.

Finally, as it can be seen from Figure~9 (lower panel), the galaxies taking part of AC114 02 are sparsely distributed
background to the main AC114 suggesting else they do not constitute a bound structure, or else the sky region it occupies
has suffered of a bad sampling.}

Note that if the relaxation occurs in two or more clusters in the line of sight, supporting evidence for this is extremely
difficult to recognize with the existing data. Czoske \etal (2001,2002) showed that the redshift
distribution of the cluster Cl0024+17 ($z=0.395$) is bimodal with a large primary component and a second foreground one,
separated by $v= 3000\kms$, suggesting that the system is undergoing a radial high velocity collision (note that both AC114
and Cl0024+17 show Butcher-Oemler effect, see the next section). The Cl0024+17 bimodality is confirmed by Zuhone \etal (2009)
using a high-resolution N-body/hydrodynamics simulation of such a collision. In the case of AC114, comparing Figure~7 with 
Figs.~1 and~3 of Czoske \etal (2002) suggests the foreground structure mentioned above as a distinct grouping although
the foreground one of Cl0024+17 is much more clearly delineated.

Following Rood \etal (1972), Kent \& Gunn (1982) and more recently Czoske \etal (2002), we have constructed the AC114
pseudo-phase diagram by plotting the redshift for each galaxy versus its projected distance from the cluster center on
Figure~10. The well-sampled part of AC114 corresponds to the very central part of the cluster Cl0024+17 extending 5~arcmin
on the Figure~2 of Czoske \etal (2002). The distribution of the galaxies in the main peak is symmetrical with respect 
to the central redshift, whereas the distribution of the foreground galaxies is evidenced at a rest frame
velocity of~$\simeq -5000 \kms$ at radii until 7~arcmin, but turns out towards smaller relative velocities to merge
with the main distribution at smaller projected distances. The background galaxies velocity at a rest frame 
of~$\simeq +4000 \kms$ are much more widely dispersed and we cannot rule out a connection with AC114; it seems more
likely that they are part of the surrounding field galaxy population. Note the ``trumpet shaped'' region extending
horizontally from the cluster center. The infall pattern around rich clusters of galaxies has been studied by Regos \&
Geller (1989) by constructing an analytic model for the distribution of galaxies around a cluster core in redshift space.
From their Figure~6-b) showing the redshift as a function of the angular separation on the sky from the cluster center,
the high-density caustic surfaces which define the infall pattern show this characteristic trumpet shape. This shape is 
interpreted by these authors as the collapsing structure in which galaxies trace the matter distribution on the scale of
the infall region and moreover, the caustics have a dependence on $\Omega_{o}$. Such model has been tested by Regos \& Geller
(1989) for the rich clusters A539, Coma, A2670, A1367 and by Reisenegger \etal (2000) for the Shapley supercluster.
Diaferio \& Geller (1997) interpreted this trumpet shape as outlining the escape velocity profile rather than caustics due
to the infall pattern. The appearance of Figure~10 supports the reality of the foreground structure as it appears to lie
mostly outside of the lower caustic. However, it is situated close to the limit of the virial radius of AC114. With a more
complete set of velocities one could perhaps resolve definitively the dynamical status of AC114 and its immediate environs.

%\begin{figure*}
%\begin{centering}
%\includegraphics[angle=-90,width=1.0\columnwidth]{vit-RA.ps}
%\includegraphics[angle=-90,width=1.0\columnwidth]{vit-Dec.ps}
%\par\end{centering}
%\caption{Velocity versus R.A. {\it left} and Dec. {\it right} of the AC114 cluster between $89000$ and $101000 \kms$. (+)
%correspond to galaxies with available velocities but no photometry while (*) are galaxies with $B_{j}-R_{f} \geq $ 2.0
%and plussed-squares are galaxies with $B_{j}-R_{f} \prec $ 2.0.} 
%\end{figure*}

\begin{figure*}
\begin{centering}
\includegraphics[width=1.2\columnwidth]{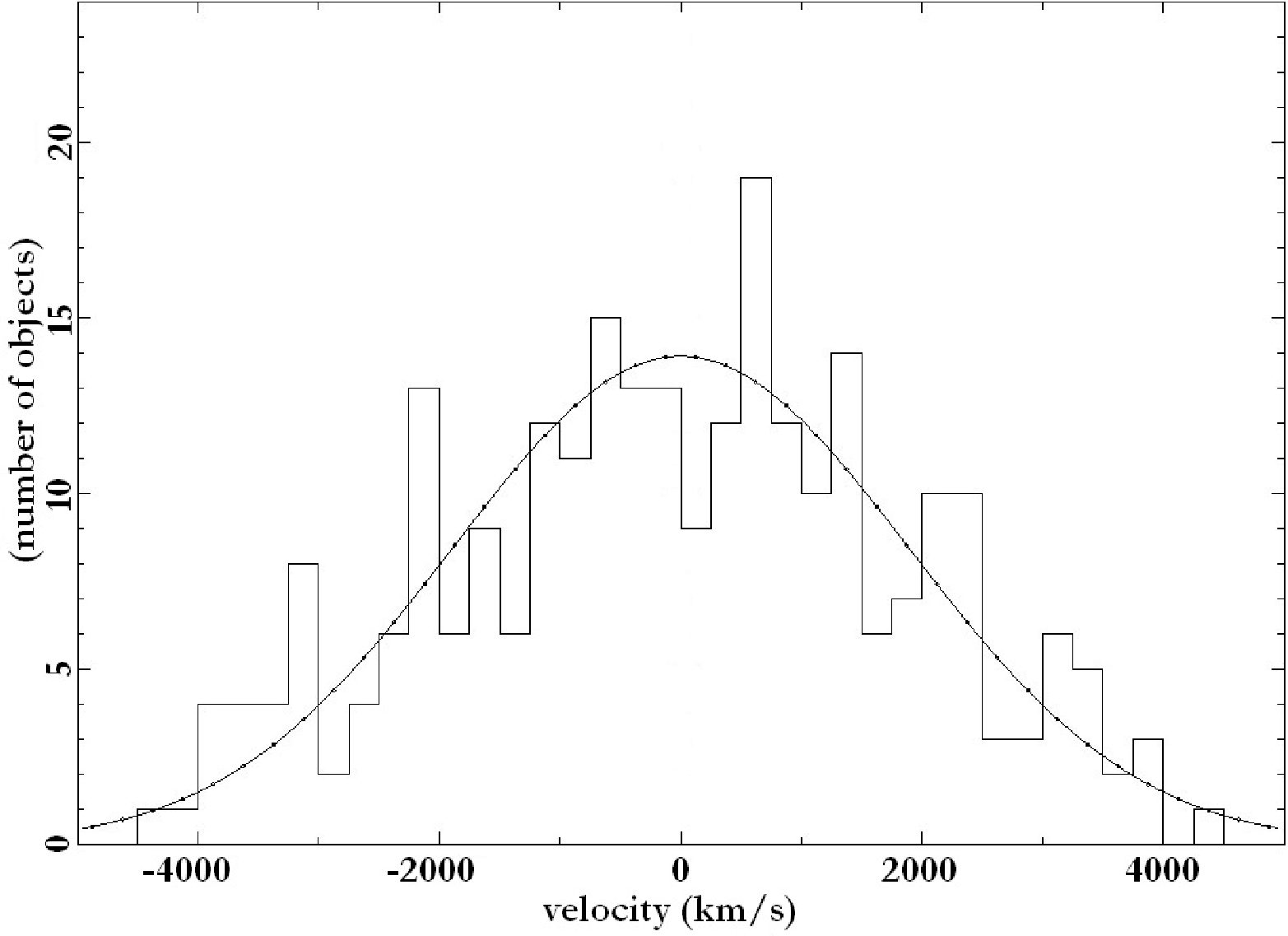}
\par\end{centering}
\caption{Rest-frame histogram of the velocity distribution of the AC114 cluster between $89000$ and $101000\kms$
with a step of $250\kms$; the gaussian with $\sigma= 1893\kms$ (rest-frame) is also shown.}
\end{figure*}

\begin{figure*}
\begin{centering}
\includegraphics[width=1.0\columnwidth]{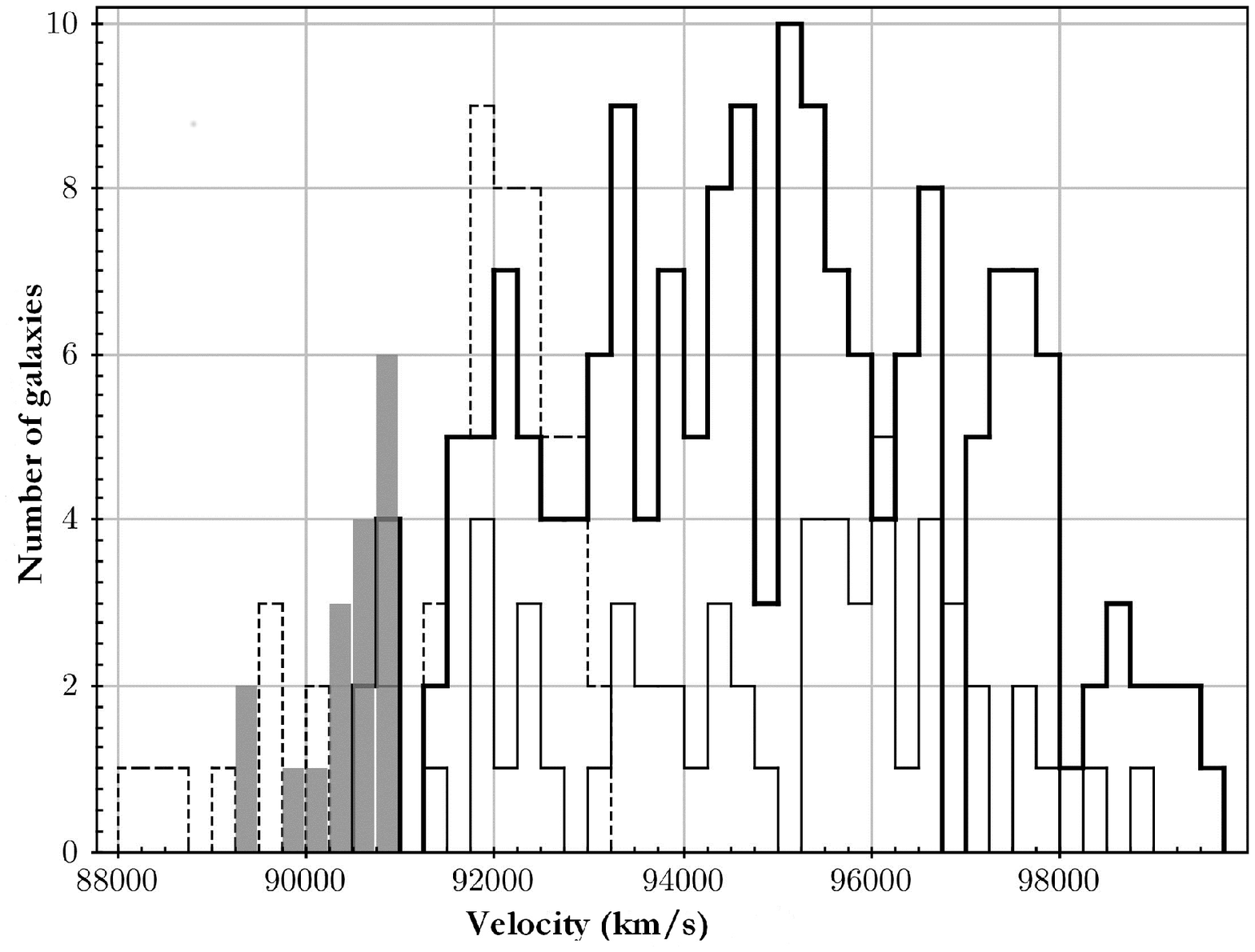}
\bigskip
\includegraphics[width=1.2\columnwidth]{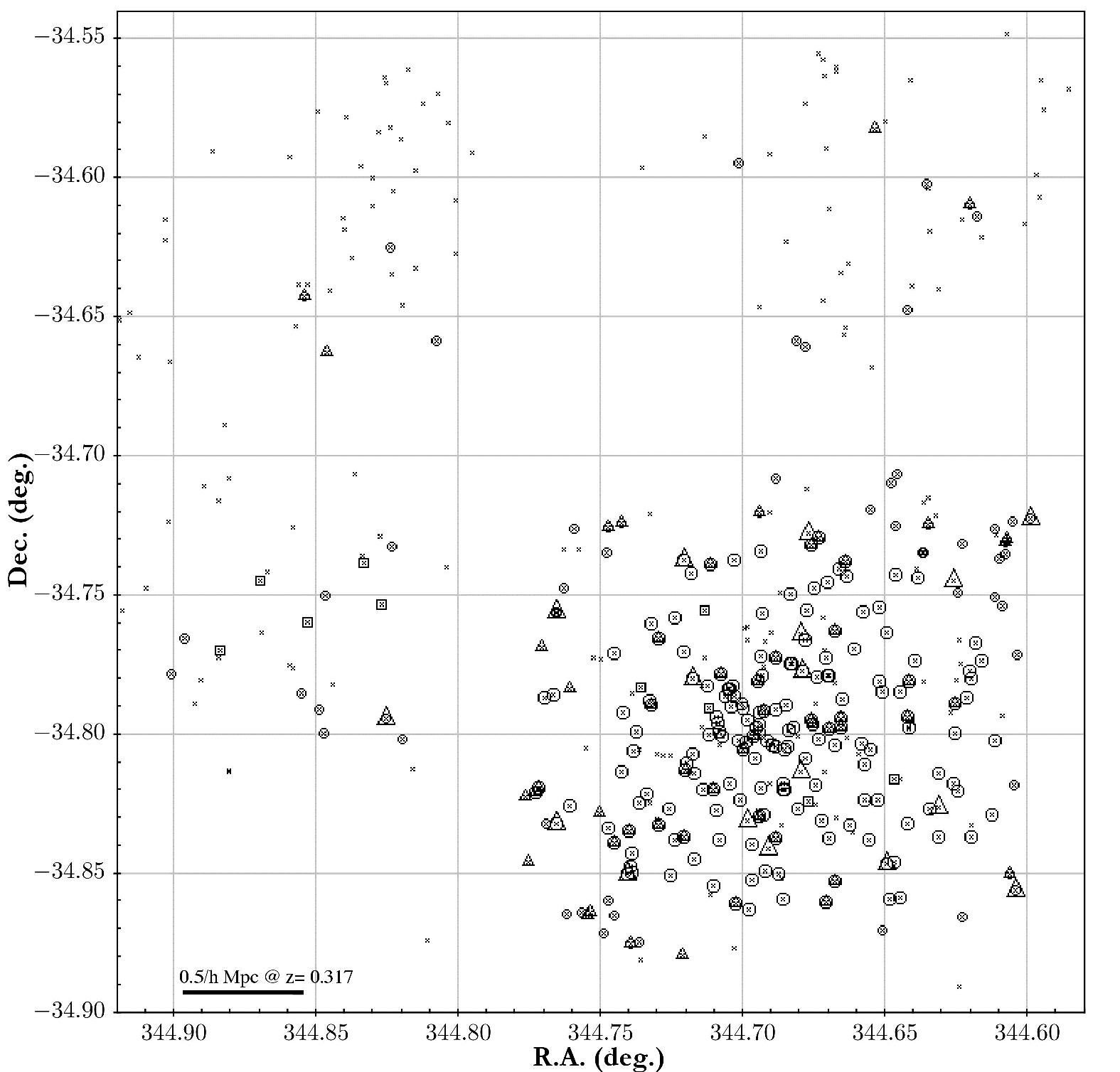}
\par\end{centering}
\caption{ {\it upper pannel}: Velocity histogram (step= $250\kms$) showing the main (heavy stepped line) and foreground
(filled rectangles) structures detected as described in the text. The {\it light stepped lines} show the distribution of
galaxies classified as interlopers for each of the structures. {\it Lower pannel}: the projected distribution of galaxies of
our spectroscopic sample ({\it small crosses}); {\it large/small open circles} show the positions of the galaxies 
belonging/interlopers to the main cluster, {\it large/small open triangles} give the positions of the
foreground/interlopers structures galaxies and {\it small/open squares} are for those belonging to the background structure
(cf. Table~3).}
\end{figure*}

\begin{figure*}
\begin{centering}
\includegraphics[width=1.3\columnwidth]{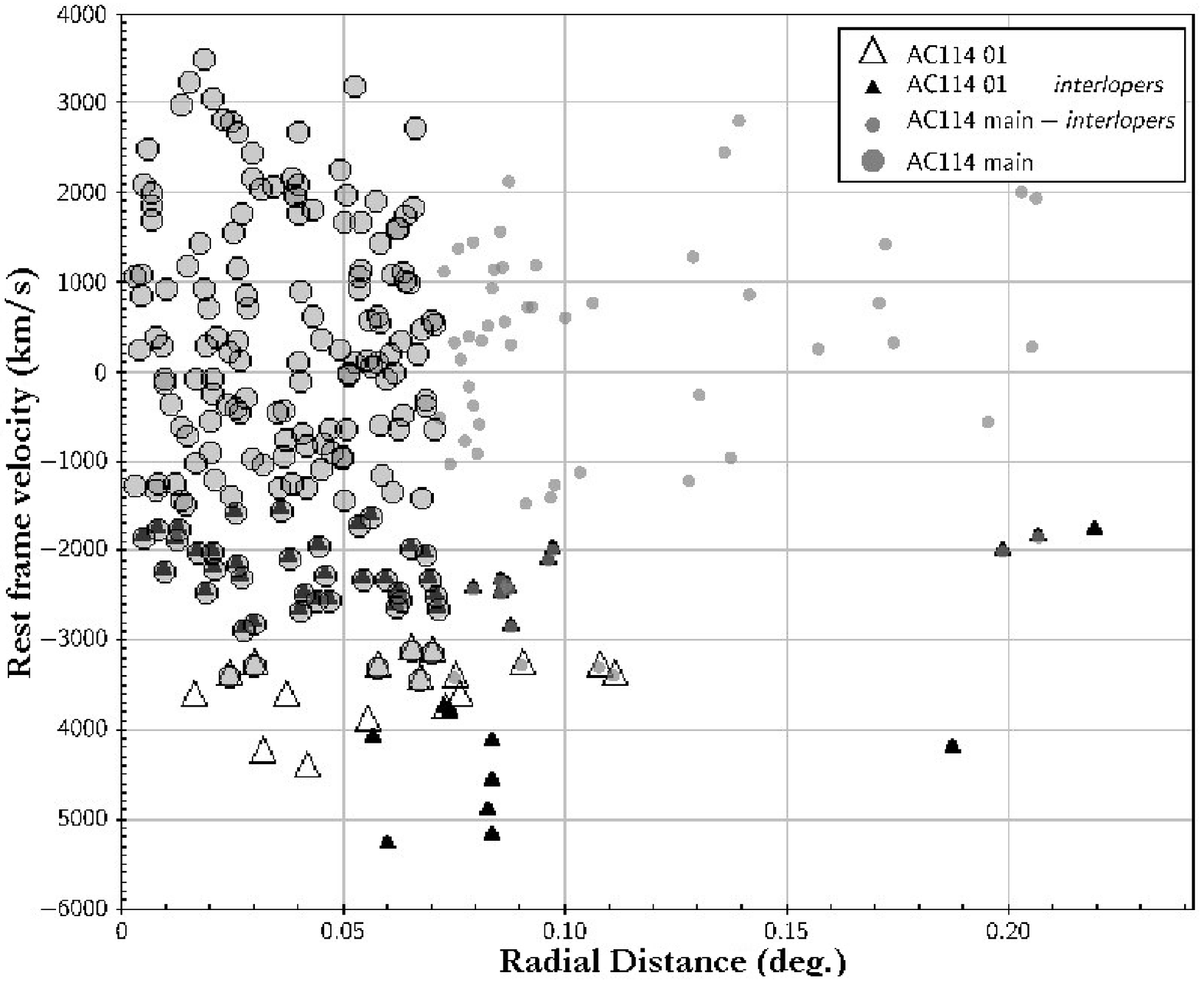}
\par\end{centering}
\caption{Relative velocity with respect to the mean cluster velocity plotted against angular distance R from the projected 
cluster center for the galaxies around the cluster velocity. The left axis expresses relative velocity with respect to the
mean velocity of the cluster; symbols are identical to Fig.9.}
\end{figure*}

We also applied the method developed by Dressler \& Shectman (1988) with the $\delta$ parameter to test for
kinematical structures defined as:

\begin{equation}
{\delta^{2}= (11/\sigma^{2}[(\overline{v}_{local} - \overline{v})^{2} + (\sigma_{local} - \sigma)^{2}]}
\end{equation}

where ${v}_{local}$ and $\sigma_{local}$ are the local velocities and dispersions calculated from the 10~nearest
neighbors of each galaxy within the $r_{200}$ radius which defines the limits of the virialized cluster from its
redshift and velocity dispersion (i.e., with average density 200 times the critical one, see e.g.,
Diaferio \etal 2001, Finn \etal 2004):

\begin{equation}
r_{200}= 1.73\frac{\sigma_{v,cl}}{1000\kms}\frac{1}{\sqrt{\Omega_{\Lambda}+\Omega_{o}(1+z_{cl})^{3}}}/h~Mpc
\end{equation}

In random redistributions of the measured galaxy velocities and positions, with $r_{200} \simeq 2.78$/h~Mpc
we found the sum of the $\delta$ values to be equal to or larger than the observed value with a
frequency of $P_{\delta}= 0.138$. This value is too marginal to conclusively verify the existence of substantial
substructures. With $\sigma_{v,cl}= 2025\kms$ Martini \etal (2007) give $r_{200} \simeq 4.25$/$h_{70}$~Mpc and
$P_{\delta}= 0.156$. Just \etal (2010) give $r_{200}$= 3.98/$h_{70}$~Mpc while Couch \etal (2001) use a value 
$r_{v}$= 3.0/$h_{100}$~Mpc.

\begin{table*}[]
\caption{Positions, photometric data and velocities for galaxies of AC114. The R column is the 
correlation peak height to the amplitude of the antisymmetric noise (see text for explanation).}
\small
\begin{tabular}{cccccccccl}
\hline
\hline
{\bf Gal.} & {\bf R.A.}    & {\bf Dec.}    & {\bf $B_{j}$}   &  {\bf $R_{f}$}   & {\bf R$_{{\rm comp}}$}  & {\bf redshift} & {\bf error} & {\bf R} & {\bf notes} \\
           & {\bf (J2000)} & {\bf (J2000)} &  {\bf mag}      & {\bf mag}       & {\bf mag} &          &             &         &          \\
\hline
Quadrant 1 &      &             &       &       &          &          &         &       & \\
  1 & 22 59 20.70 & -34 42 24.6 &       &       &    20.0  & 0.39852  & 0.00030 &  3.50 & em:{H$\alpha$},S1 \\
  2 & 22 59 31.37 & -34 42 28.4 &       &       &    20.5  & 0.35306  & 0.00018 &       & em:OII,{H$\beta$},2OIII,{H$\alpha$},S1 \\
  3 & 22 59 33.46 & -34 42 39.8 &       &       &    18.0  & 0.05717  & 0.00011 &       & em:{H$\beta$},2OIII,{H$\alpha$},S1 \\
  5 & 22 59 32.31 & -34 42 58.2 & 20.2  &       &    20.1  & 0.05842  & 0.00009 &       & em:{H$\beta$},2OIII,{H$\alpha$},S1 \\
  8 & 22 59 36.51 & -34 43 25.4 & 20.2  &       &    20.3  & 0.38058  & 0.00022 &       & em:{H$\alpha$},S1 \\
  9 & 22 59 25.99 & -34 43 31.9 &       &       &          & 0.21032? &         &       & very uncertain \\
 10 & 22 59 18.60 & -34 43 43.6 & 20.5  & 18.8  &    18.9  & 0.21940  & 0.00024 &  4.76 & em:{H$\alpha$},S1 65840$\kms$ \\
 11 & 22 59 17.61 & -34 43 57.1 &       &       &    19.3  & 0.31185  & 0.00019 &  3.27 & \\
 12 & 22 59 20.05 & -34 44 09.4 & 20.8  & 19.5  &    18.2  & 0.22127  & 0.00012 &  3.56 & em:{H$\beta$},{H$\alpha$} 66509$\kms$ \\
 13 & 22 59 20.00 & -34 44 20.3 &       &       &    20.0  & 0.40980  & 0.00028 &  4.48 & \\
 15 & 22 59 28.09 & -34 44 29.0 &       &       &    19.4  & 0.07309  & 0.00045 &       & em:{H$\beta$},2OIII,{H$\alpha$} \\
 16 & 22 59 28.82 & -34 44 44.1 &       &       &    18.6  & 0.40823  & 0.00029 &  3.61 & \\
 17 & 22 59 38.42 & -34 44 52.4 &       &       &    20.5  & 0.34606  & 0.00029 &  3.04 & very weak \\
 18 & 22 59 23.15 & -34 45 03.0 &       &       &    19.7  & 0.31301  & 0.00007 &       & em:{H$\alpha$},S1 \\
 19 & 22 59 18.51 & -34 45 13.1 & 21.2  & 20.3  &    19.6  & 0.40876  & 0.00022 &       & em:OII,{H$\beta$},2OIII,{H$\alpha$},S1 \\ 
 20 & 22 59 40.37 & -34 45 21.8 &       &       &    18.1  & 0.23192  & 0.00018 &  5.23 & \\
 22 & 22 59 24.80 & -34 45 36.6 & 22.3  & 19.6  &    18.8  & 0.41015  & 0.00025 &  5.19 & em:{H$\beta$},{H$\alpha$} \\
 24 & 22 59 28.59 & -34 45 49.8 & 21.6  & 19.6  &    18.9  & 0.33369  & 0.00018 &  4.84 & em:OII \\
 25 & 22 59 35.08 & -34 45 57.0 & 22.4  &       &    20.9  & 0.32344  & 0.00028 &  3.12 & \\
 26 & 22 59 32.09 & -34 46 13.6 &       &       &    19.6  & 0.40926  & 0.00027 &  3.65 & \\
 27 & 22 59 32.34 & -34 46 23.3 &       &       &    19.9  & 0.46615  & 0.00022 &  3.22 & \\
 28 & 22 59 26.16 & -34 46 32.5 & 20.2  & 19.2  &    19.0  & 0.35118  & 0.00028 &  3.11 & weak \\
 30 & 22 59 36.18 & -34 46 43.1 & 20.8  & 18.8  &    19.6  & 0.31860  & 0.00033 &  3.15 & \\
 31 & 22 59 33.65 & -34 46 49.2 &       &       &    19.9  & 0.33342  &         &       & measured on {H$\alpha$} \\
 34 & 22 59 25.25 & -34 47 05.9 &       &       &    20.0  & 0.32794  & 0.00042 &  3.02 & very weak \\
 36 & 22 59 34.22 & -34 47 19.9 &       &       &    19.6  & 0.33300  & 0.00027 &  3.12 & very weak \\
 37 & 22 59 23.76 & -34 47 27.4 & 21.0  & 19.8  &    19.6  & 0.31604  & 0.00004 &       & em:{H$\beta$},2OIII,{H$\alpha$},S1 \\
 38 & 22 59 18.07 & -34 47 39.8 &       &       &    19.9  & 0.30236  & 0.00016 &  3.10 & \\
 40 & 22 59 23.41 & -34 47 57.6 & 20.6  & 18.9  &    18.1  & 0.32283  & 0.00027 &  3.37 & em:{H$\beta$},2OIII,{H$\alpha$},S1: 96857$\kms$ \\
 41 & 22 59 16.77 & -34 48 06.8 &       &       &    20.0  & 0.32061  & 0.00045 &       & \\
 44 & 22 59 15.92 & -34 48 43.8 & 22.2  & 21.1  &    19.1  & 0.28628? & 0.00060 &  4.04 & uncertain \\
45a & 22 59 31.40 & -34 48 50.3 &       &       &          & 0.35159  &         &       & measured on {H$\alpha$},S1 \\
45b & 22 59 31.40 & -34 48 50.4 &       &       &          & 0.34461  &         &       & measured on {H$\alpha$},S1 \\
Quadrant 2 &      &             &       &       &             &          &         &       & \\
 1a & 22 59 16.30 & -34 33 40.6 & 20.8  & 19.0  &    19.6  & 0.25782  & 0.00023 &  3.43 & em:{H$\beta$},2OIII,{H$\alpha$} \\
 1b & 22 59 16.30 & -34 33 40.6 &       &       &             & 0.33069  & 0.00024 &  3.18 & em:{H$\alpha$}, 2 galaxies \\
  2 & 22 59 18.23 & -34 33 50.7 &       &       &    18.2  & 0.49578  & 0.00029 &       & em:OII,{H$\beta$},2OIII,{H$\alpha$},S1 \\
  3 & 22 59 18.00 & -34 33 58.5 &       &       &    20.5  & 0.41000  & 0.00033 &       & em:OII,{H$\beta$},2OIII,{H$\alpha$} \\ 
  4 & 22 59 13.67 & -34 34 11.1 &       &       &    19.3  & 0.40846  & 0.00033 &       & em:{H$\beta$},2OIII \\
  5 & 22 59 15.03 & -34 34 23.5 & 22.5  & 20.1  &    18.8  & 0.41440  & 0.00034 &  3.01 & very uncertain \\
  6 & 22 59 23.91 & -34 34 32.7 & 21.3  & 19.2  &    18.9  & 0.42284  & 0.00021 &  6.45 & \\
  7 & 22 59 21.45 & -34 34 41.8 &       &       &    18.6  & 0.23156  & 0.00024 &  3.36 & very weak \\
  8 & 22 59 12.89 & -34 34 48.4 &       &       &    21.2  & 0.44924  & 0.00018 &  3.03 & weak \\
  9 & 22 59 17.73 & -34 34 54.1 &       &       &    20.9  & 0.31359? &         &       & very uncertain \\
 10 & 22 59 18.77 & -34 35 02.2 & 21.5  & 19.9  &    19.5  & 0.40848  & 0.00023 & 3.36  & \\
 11 & 22 59 16.88 & -34 35 08.5 &       &       &    17.6  & 0.34546  &         &       & measured on {H$\alpha$} \\
 12 & 22 59 32.84 & -34 35 24.7 & 21.9  &       &    19.2  & 0.34646  & 0.00021 &  3.08 & em:{H$\alpha$} \\
 13 & 22 59 26.31 & -34 35 32.2 &       &       &    17.8  & 0.27013  & 0.00038 &  3.52 & very weak \\
 14 & 22 59 20.21 & -34 35 43.7 &       &       &    18.7  & 0.37370  & 0.00009 &       & em:OII,{H$\alpha$},S1 \\
 15 & 22 59 15.61 & -34 35 51.1 &       &       &    20.7  & 0.41045  & 0.00026 &  4.01 & \\
 16 & 22 59 19.18 & -34 35 59.7 & 21.4  & 20.4  &    19.7  & 0.34469  & 0.00021 & 3.41  & \\
 18 & 22 59 12.16 & -34 36 28.0 & 21.0  &       &    19.4  & 0.38110  & 0.00042 &  4.03 & \\
 19 & 22 59 19.31 & -34 36 38.6 & 22.00 & 20.1  &    18.8  & 0.37706  & 0.00021 & 6.71  & \\
 21 & 22 59 21.69 & -34 36 53.5 &       &       &    19.2  & 0.41304  & 0.00020 &  6.52 & \\
 22 & 22 59 36.79 & -34 36 53.5 &       &       &    21.3  & 0.39104  &         &       & em:{H$\alpha$} \\
 23 & 22 59 21.57 & -34 37 07.8 & 19.3  & 18.4  &    18.3  & 0.41310  & 0.00022 &  6.81 & \\
 24 & 22 59 36.81 & -34 37 21.2 &       &       &    19.2  & 0.37972  & 0.00052 &       & em:OII,{H$\beta$},2OIII \\
\hline
\end{tabular}
\end{table*}
\begin{table*}[]
\begin{tabular}{cccccccccl}
\hline
\hline
{\bf Gal.} & {\bf R.A.}    & {\bf Dec.}    & {\bf  $B_{j}$}   &  {\bf $R_{f}$}  &{\bf R$_{{\rm comp}}$} & {\bf redshift} & {\bf error} & {\bf R} & {\bf notes} \\
           & {\bf (J2000)} & {\bf (J2000)} & {\bf mag} & {\bf mag} & {\bf mag} &              &             &         &          \\
\hline
 26 & 22 59 12.34 & -34 37 37.5 & 22.3  & 20.2  & 19.4  & 0.48676  & 0.00018 &       & em:OII,{H$\beta$},2OIII \\
 27 & 22 59 21.00 & -34 37 46.0 & 20.6  & 20.0  & 19.9  & 0.49642  & 0.00024 & 4.53  & \\
 28 & 22 59 15.56 & -34 37 57.2 & 20.8  & 19.4  & 19.7  & 0.47064  & 0.00027 &  3.16 & \\
 29 & 22 59 17.62 & -34 38 03.8 &       &       & 20.9  & 0.39781  & 0.00055 &  3.06 & very weak, uncertain \\
 31 & 22 59 24.67 & -34 38 18.4 & 21.4  & 20.6  & 18.5  & 0.49717  & 0.00016 &       & em:OII,{H$\beta$},2OIII,{H$\alpha$} \\
 32 & 22 59 25.00 & -34 38 35.6 & 20.8  & 19.5  & 19.5  & 0.30913  & 0.00028 &       & em:{H$\beta$},2OIII,{H$\alpha$} \\
 33 & 22 59 16.65 & -34 38 43.7 & 20.1  & 18.1  & 20.4  & 0.46922  & 0.00051 &  3.17 & em:{H$\beta$} 140472$\kms$ \\
 35 & 22 59 40.57 & -34 39 05.1 &       &       & 19.7  & 0.31290  & 0.00016 &  7.72 & \\
 36 & 22 59 25.79 & -34 39 11.8 & 21.4  & 20.0  & 19.8  & 0.46367  & 0.00020 &  6.71 & \\
 37 & 22 59 13.94 & -34 39 33.0 & 21.8  & 20.7  & 18.7  & 0.32064  & 0.00016 &  3.42 & \\
 38 & 22 59 23.08 & -34 39 47.1 &       &       & 20.1  & 0.29869  & 0.00017 &       & em:{H$\beta$},2OIII,{H$\alpha$},S1 \\
 39 & 22 59 39.01 & -34 39 53.1 &       &       & 20.6  & 0.37203  & 0.00012 &       & em:OII,{H$\beta$},2OIII,{H$\alpha$} \\
 40 & 22 59 36.39 & -34 39 58.4 &       &       & 20.3  & 0.41913  & 0.00023 &  4.96 & \\
 41 & 22 59 43.80 & -34 40 06.7 &       &       & 20.5  & 0.31907  & 0.00026 &  3.08 & \\
Quadrant 3 &      &             &       &       &       &          &         &       & \\
  1 & 22 58 41.57 & -34 33 18.1 & 20.3  & 18.8  & 19.8  & 0.32951  & 0.00015 &       & em:{H$\beta$},2OIII,{H$\alpha$},S1 \\
  2 & 22 58 41.21 & -34 33 27.8 & 20.5  & 18.8  & 18.7  & 0.33307  & 0.00011 &       & em:{H$\beta$},2OIII,{H$\alpha$},S1 \\
  3 & 22 58 40.11 & -34 33 34.6 & 21.9  & 19.2  & 19.4  & 0.33127  & 0.00024 &  5.48 & \\
  4 & 22 58 40.08 & -34 33 42.2 & 21.3  & 19.9  & 20.0  & 0.21259  & 0.00019 &       & em:{H$\beta$},2OIII,{H$\alpha$},S1 \\
  5 & 22 58 41.10 & -34 33 49.0 &       &       & 20.5  & 0.33189  & 0.00020 &       & em:{H$\beta$},2OIII,{H$\alpha$},S1 \\
  6 & 22 58 33.87 & -34 33 53.5 &       &       & 19.2  & 0.38938? &         &       & uncertain \\
  8 & 22 58 42.73 & -34 34 23.0 & 22.0  &       & 19.7  & 0.39969  & 0.00028 &  3.48 & em:{H$\alpha$}:119724$\kms$ \\
 10 & 22 58 22.64 & -34 34 33.3 & 19.8  & 18.7  & 18.8  & 0.31365  & 0.00012 &  4.63 & \\
 11 & 22 58 35.95 & -34 34 45.6 & 22.4  & 20.6  & 20.1  & 0.21245  & 0.00026 &       & em:{H$\beta$},2OIII,{H$\alpha$},S1 \\
 12 & 22 58 36.88 & -34 34 58.2 & 21.5  & 19.6  & 19.0  & 0.30939  & 0.00009 &       & em:{H$\beta$},2OIII,{H$\alpha$},S1 \\
 13 & 22 58 51.29 & -34 35 07.6 &       &       & 19.3  & 0.36347  & 0.00044 &  3.02 & very weak \\
 15 & 22 58 41.00 & -34 35 21.4 &       &       & 20.3  & 0.47562  & 0.00038 &  3.01 & very weak \\
 16 & 22 58 45.84 & -34 35 30.4 & 20.9  & 20.4  & 20.0  & 0.17739  & 0.00012 &       & em:{H$\beta$},2OIII,{H$\alpha$},S1 \\
 17 & 22 58 48.43 & -34 35 39.7 &       &       & 20.1  & 0.31846  & 0.00041 &  3.22 & \\
 18 & 22 58 56.50 & -34 35 48.2 & 20.8  & 19.3  & 19.5  & 0.22176  & 0.00009 &       & em:{H$\beta$},2OIII,{H$\alpha$},S1 \\
 19 & 22 58 23.26 & -34 35 57.0 & 20.0  & 18.6  & 18.9  & 0.15860  & 0.00015 &  4.84 & \\
 20 & 22 58 32.46 & -34 36 07.9 &       &       & 19.2  & 0.32601  & 0.00029 &  3.01 & very weak \\
 21 & 22 58 32.39 & -34 36 15.5 & 21.1  & 19.2  & 20.8  & 0.35231  & 0.00021 &       & em:{H$\beta$},{H$\alpha$},S1,S2 \\
 23 & 22 58 22.99 & -34 36 26.7 & 22.1  & 20.5  & 20.5  & 0.43221  & 0.00036 &  3.39 & \\
 24 & 22 58 28.94 & -34 36 34.7 &       &       & 18.2  & 0.30386  & 0.00045 &  3.01 & very weak \\
 25 & 22 58 40.69 & -34 36 41.1 &       &       & 20.1  & 0.47190  & 0.00039 &  3.15 & \\
 26 & 22 58 28.24 & -34 36 48.7 &       &       & 20.2  & 0.31477  & 0.00024 &  4.07 & \\
 27 & 22 58 29.50 & -34 36 55.9 &       &       & 19.6  & 0.41028? &         &       & uncertain \\
 28 & 22 58 24.28 & -34 37 02.1 &       &       & 18.4  & 0.25761  & 0.00018 &  5.51 & \\
 29 & 22 58 32.32 & -34 37 07.5 & 21.6  & 20.1  & 20.6  & 0.47283  & 0.00038 &  3.15 & \\
 30 & 22 58 27.93 & -34 37 16.9 & 21.2  & 20.1  & 21.1  & 0.40221  & 0.00023 &       & em:OII, {H$\beta$},2OIII,{H$\alpha$} \\  
 31 & 22 58 44.38 & -34 37 24.2 & 22.2  & 20.2  & 20.1  & 0.39764  & 0.00040 &  3.04 & em:{H$\beta$},{H$\alpha$} \\
 33 & 22 58 39.09 & -34 37 52.3 &       &       & 18.8  & 0.34480  &         &       & measured on {H$\alpha$} \\
 34 & 22 58 39.66 & -34 38 03.8 &       &       & 20.1  & 0.35179  & 0.00011 &       & em:OII,{H$\beta$},2OIII,{H$\alpha$},S1 \\
 35 & 22 58 33.67 & -34 38 18.9 & 21.4  & 20.1  & 18.9  & 0.43091  & 0.00014 &       & em:{H$\beta$},2OIII,{H$\alpha$},S1 \\
 36 & 22 58 31.51 & -34 38 24.2 & 22.3  & 20.7  & 20.5  & 0.20326  & 0.00031 &       & em:{H$\alpha$},S1 \\
 38 & 22 58 41.21 & -34 38 40.5 &       &       & 19.3  & 0.43205  & 0.00029 &  3.58 & em: OII \\
 39 & 22 58 46.63 & -34 38 45.7 & 21.9  & 20.6  & 20.2  & 0.39935  & 0.00011 &       & em:OII,{H$\beta$},2OIII,{H$\alpha$},S1 \\
 40 & 22 58 34.07 & -34 38 53.1 & 21.9  & 19.9  & 19.3  & 0.31828  & 0.00031 &  4.12 & \\
 43 & 22 58 39.42 & -34 39 13.5 &       &       & 20.2  & 0.27849  & 0.00033 &  3.11 & \\
 44 & 22 58 39.45 & -34 39 24.1 & 22.2  & 19.4  & 19.5  & 0.22464  & 0.00017 &  3.34 & em: {H$\alpha$},S1 67344$\kms$ \\
 45 & 22 58 43.50 & -34 39 32.3 & 21.6  & 20.6  & 20.1  & 0.32099  & 0.00020 &  3.07 & \\
 46 & 22 58 42.70 & -34 39 40.6 &       &       & 20.4  & 0.32952  & 0.00034 &  3.02 & very weak \\
 49 & 22 58 37.09 & -34 40 06.8 & 19.4  & 18.2  & 18.9 & 0.20657  & 0.00019 &  4.82 & rm:{H$\alpha$},S1 61753$\kms$ \\
Quadrant 4 &      &             &       &       &       &          &         &       & \\
  2 & 22 58 34.99 & -34 42 24.0 &       &       & 19.5  & 0.31982  & 0.00035 &  5.82 & em:{H$\alpha$} weak \\
  3 & 22 58 45.17 & -34 42 28.9 & 22.6  & 19.1  & 20.1  & 0.32038  & 0.00021 &  3.18 & \\
  4 & 22 58 35.53 & -34 42 35.8 &       &       & 19.7  & 0.31108  & 0.00014 &  3.42 & \\
\hline
\end{tabular}
\end{table*}
\begin{table*}[]
\begin{tabular}{cccccccccl}
\hline
\hline
{\bf Gal.} & {\bf R.A.}    & {\bf Dec.}    & {\bf $B_{j}$}   &  {\bf $R_{f}$}   & {\bf R$_{{\rm comp}}$}  & {\bf redshift} & {\bf error} & {\bf R} & {\bf notes} \\
           & {\bf (J2000)} & {\bf (J2000)} & {\bf mag} & {\bf mag} & {\bf mag} &           &             &         &          \\
\hline
  5 & 22 58 42.57 & -34 42 42.5 &       &       & 20.6  & 0.42558  & 0.00012 &       & em:{H$\beta$},2OIII,{H$\alpha$},S1 \\
  7 & 22 58 32.37 & -34 42 55.1 & 20.4  & 19.7  & 20.0  & 0.16952  & 0.00009 &       & em:{H$\beta$},2OIII,{H$\alpha$},S1 \\
  8 & 22 58 32.75 & -34 43 02.0 &       &       & 19.5  & 0.17061  & 0.00022 &       & em:{H$\beta$},2OIII,{H$\alpha$},S1 \\
 9a & 22 58 31.70 & -34 43 18.8 & 21.0  & 18.9  & 18.4  & 0.29095  & 0.00009 & 10.95 & \\
 9b & 22 58 31.70 & -34 43 18.8 &       &       &       & 0.25733  &         &       & em:{H$\alpha$} \\
 10 & 22 58 25.31 & -34 43 25.5 & 20.5  & 18.8  & 19.5  & 0.31228  &         &       & em:{H$\alpha$} very weak \\
 11 & 22 58 32.43 & -34 43 30.3 & 19.0  & 17.4  & 20.0  & 0.30473  & 0.00008 &  3.02 & very weak \\
 12 & 22 58 26.84 & -34 43 34.1 & 22.1  & 21.0  & 19.7  & 0.31166  & 0.00026 &  3.05 & \\
 13 & 22 58 26.64 & -34 43 44.0 & 22.4  & 20.5  & 20.9  & 0.47564  & 0.00016 &  3.40 & \\
 14 & 22 58 41.60 & -34 43 46.5 & 20.5  & 18.9  & 20.7  & 0.30460  & 0.00007 &       & em:OII,{H$\beta$},2OIII,{H$\alpha$},S1 \\
 15 & 22 58 25.91 & -34 43 52.0 & 21.2  & 19.8  & 20.9  & 0.30799  & 0.00030 &       & em:{H$\beta$},2OIII \\
 16 & 22 58 29.50 & -34 43 55.5 & 22.2  & 19.8  & 20.4  & 0.31854  & 0.00035 &  3.52 & \\
 18 & 22 58 25.81 & -34 44 06.4 &       &       & 20.6  & 0.32247  & 0.00029 &  3.19 & \\
 19 & 22 58 26.24 & -34 44 12.1 &       &       & 20.2  & 0.31070  & 0.00011 &       & em:{H$\beta$},2OIII,{H$\alpha$},S1 \\
 20 & 22 58 39.44 & -34 44 18.6 &       &       & 19.6  & 0.30863  & 0.00011 &       & em:{H$\beta$},2OIII,{H$\alpha$},S1 \\
 21 & 22 58 50.72 & -34 44 16.7 &       &       & 19.5  & 0.30605  & 0.00024 &  4.32 & \\
 22 & 22 58 39.87 & -34 44 28.0 &       &       & 19.4  & 0.31445  & 0.00022 &  5.69 & \\
 23 & 22 58 35.11 & -34 44 34.8 & 21.9  &       & 20.6  & 0.32268  & 0.00030 &  3.09 & \\
 24 & 22 58 33.19 & -34 44 39.7 &       &       & 20.5  & 0.31448  & 0.00028 &  3.12 & \\
 25 & 22 58 30.22 & -34 44 42.4 &       &       & 19.4  & 0.30132  & 0.00035 &  3.20 & \\
 26 & 22 58 42.00 & -34 44 51.2 & 20.6  & 18.9  & 18.9  & 0.32462  & 0.00012 &       & em:OII,{H$\beta$},2OIII,{H$\alpha$},S1 \\
 27 & 22 58 29.86 & -34 44 55.5 &       &       & 21.4  & 0.31265  & 0.00023 &  3.08 & \\
 28 & 22 58 26.83 & -34 45 02.6 &       &       & 20.0  & 0.31467  & 0.00032 &  3.53 & \\
 31 & 22 58 36.47 & -34 45 17.4 & 21.0  & 19.3  & 19.1  & 0.31779  & 0.00024 &  5.33 & \\
 32 & 22 58 38.00 & -34 45 24.1 & 20.2  & 17.7  & 18.2  & 0.31708  & 0.00024 &  5.65 & \\
 33 & 22 58 41.30 & -34 45 30.6 & 21.1  & 19.6  & 19.8  & 0.10193  & 0.00029 &       & em:OII,{H$\beta$},2OIII,{H$\alpha$},S1 \\
 35 & 22 58 47.92 & -34 45 45.0 & 21.7  & 19.5  & 20.2  & 0.17040  & 0.00011 &       & em:{H$\beta$},2OIII,{H$\alpha$},S1 \\
 36 & 22 58 35.91 & -34 45 50.1 & 19.8  & 17.7  & 18.4  & 0.31307  & 0.00019 &  5.65 & \\
 37 & 22 58 29.65 & -34 45 57.5 &       &       & 19.1  & 0.17051  & 0.00018 &  3.01 & uncertain \\
 38 & 22 58 28.42 & -34 46 02.3 & 22.0  & 19.8  & 19.2  & 0.30100  & 0.00026 &  5.63 & \\
 39 & 22 58 38.57 & -34 46 10.7 &       &       & 20.1  & 0.32594  & 0.00047 &  3.43 & \\
 40 & 22 58 24.92 & -34 46 18.7 & 20.8  & 19.0  & 19.0  & 0.31779  & 0.00004 &       & em:{H$\alpha$} \\
 41 & 22 58 27.85 & -34 46 26.3 &       &       & 19.7  & 0.31816  & 0.00033 &  4.55 & \\
 42 & 22 58 29.75 & -34 46 32.1 & 22.2  & 20.5  & 19.9  & 0.50228  & 0.00035 &  3.04 & \\
 43 & 22 58 28.88 & -34 46 40.3 &       &       & 20.2  & 0.31717  & 0.00033 &  3.68 & \\
 44 & 22 58 40.93 & -34 46 45.7 & 20.6  & 18.6  & 18.3  & 0.31523  & 0.00030 &  4.73 & \\
 45 & 22 58 36.47 & -34 46 51.2 &       &       & 20.6  & 0.31389  & 0.00019 &  4.66 & \\
 46 & 22 58 51.00 & -34 47 00.6 & 20.8  & 19.5  & 20.1  & 0.32953  & 0.00028 &  3.10 & \\
 47 & 22 58 34.80 & -34 47 07.1 & 22.2  &       & 19.3  & 0.31674  & 0.00019 &  6.24 & \\
 48 & 22 58 29.14 & -34 47 13.1 & 21.6  & 20.0  & 20.0  & 0.31462  & 0.00022 &  5.29 & \\
 49 & 22 58 39.56 & -34 47 17.4 & 21.8  & 19.1  & 20.0  & 0.31818  & 0.00024 &  4.89 & \\ 
 50 & 22 58 48.11 & -34 47 21.3 &       &       & 20.2  & 0.33032  & 0.00039 &  3.02 & very weak \\
 51 & 22 58 46.31 & -34 47 31.5 & 20.8  & 18.8  & 19.1  & 0.30954  & 0.00019 &  6.32 & \\
 52 & 22 58 26.11 & -34 47 37.5 & 22.6  & 20.9  & 18.2  & 0.38782  & 0.00037 &  3.04 & very weak \\
 54 & 22 58 33.80 & -34 47 46.6 & 22.3  & 19.5  & 18.9  & 0.34779  & 0.00030 &  3.06 & weak \\
55a & 22 58 33.96 & -34 47 53.3 &       &       & 20.2  & 0.31776  & 0.00031 &  3.26 & \\
55b & 22 58 33.96 & -34 47 53.3 &       &       &       & 0.17199  &         &       & em:{H$\beta$},2OIII,{H$\alpha$},S1 \\
 56 & 22 58 44.05 & -34 47 59.4 &       &       & 18.4  & 0.32814  & 0.00044 &  3.31 & \\
 57 & 22 58 49.74 & -34 48 01.9 & 21.6  & 19.7  & 19.1  & 0.31451  & 0.00028 &  3.35 & \\
 58 & 22 58 41.65 & -34 48 06.7 & 21.9  & 19.2  & 19.2  & 0.31082  & 0.00028 &  6.10 & \\
 59 & 22 58 45.48 & -34 48 14.9 &       &       & 20.1  & 0.31831  & 0.00035 &  3.04 & weak \\
 60 & 22 58 44.45 & -34 48 21.3 &       &       & 19.0  & 0.32473  & 0.00031 &  4.42 & \\
 61 & 22 58 38.22 & -34 48 27.8 & 20.7  & 18.6  & 19.7  & 0.72072  & 0.00014 &  3.05 & \\
 62 & 22 58 42.81 & -34 48 33.4 & 21.0  & 18.9  & 29.9  & 0.31090  & 0.00047 &  5.43 & \\
 63 & 22 58 52.67 & -34 48 39.1 &       &       & 20.2  & 0.31874  & 0.00024 &  4.98 & \\
 64 & 22 58 31.46 & -34 48 53.3 &       &       & 18.3  & 0.31447  & 0.00030 &  5.28 & \\
 65 & 22 58 34.69 & -34 49 01.3 &       &       & 20.6 & 0.47788  & 0.00023 &  5.33 \\   
\hline
\end{tabular}
\bigskip
\end{table*}

\section{Galaxy color distribution and the Butcher-Oemler effect}\label{Color distribution}

As UK-J $B_{j}$ and ESO-R or POSS-I-E $R_{f}$ magnitudes are available from superCOSMOS (Maddox \etal 1990a, 1990b)
a color-magnitude diagram $B_{j}-R_{f}$ versus $R_{f}$ for AC114 member galaxies identified from their redshift is shown
in Figure~11 (fore - and background objects have been removed). Note that plusses and crosses correspond respectively
to absorption and emission lines galaxies observed by Couch \& Sharples (1987), while squares and stars correspond
to absorption and emission lines galaxies of the present work and from various literature sources listed in NED,
including the 6~AGN from Martini \etal (2006). The excess of blue galaxy cluster members ranging from 16 to 30\%
is known as the Butcher-Oemler (BO) effect (Butcher \& Oemler 1978). As discussed by Couch \& Sharples (1987),
a ``red'' galaxy sequence is defined with objects having $B_{j}-R_{f} \geq 2.0$. We identify this galaxy sequence
extending to $R_{f}= 21.5$ distributed along the regression line on the top of the Figure~11 with
$B_{j}-R_{f} = -0.0435B_{j} + 3.169$. This sequence is well separated from the ``blue'' galaxy area, where all
emission-lines galaxies are present. The ratio of emission/absorption lines galaxies belonging to AC114 is
19\% in Couch \& Sharples (1987) and 24\% in the present work. No obvious dynamical difference is seen between
the red and blue populations as the red one with 47 objects gives
$\overline{z}= 0.31757$ with ${\sigma}= 2019^{+115}_{-103}\kms$ and the blue one with 37~objects gives
$\overline{z}= 0.31723$ with ${\sigma}= 1835^{+108}_{-121}\kms$.

As quoted in Couch \& Sharples (1987) BO cluster member galaxies represent 36\% of the total population in AC114,
while Martini \etal (2007) found a ratio of 26\% brighter than $M_{R}=-20$ and we have 44\% of blue objects in the
present work. Such a discrepancy is a consequence of our galaxy sample, where most of them are located in the central
region of AC114. It is also an effect of the larger number of galaxy cluster members with available photometry as well
as the spectroscopy limiting magnitude of our sample which reaches magnitudes as faint as $R_{f}$= 21.1 (i.e., more than
one magnitude fainter than the previous studies).

\begin{figure*}
\begin{centering}
\includegraphics[angle=-90,width=1.3\columnwidth]{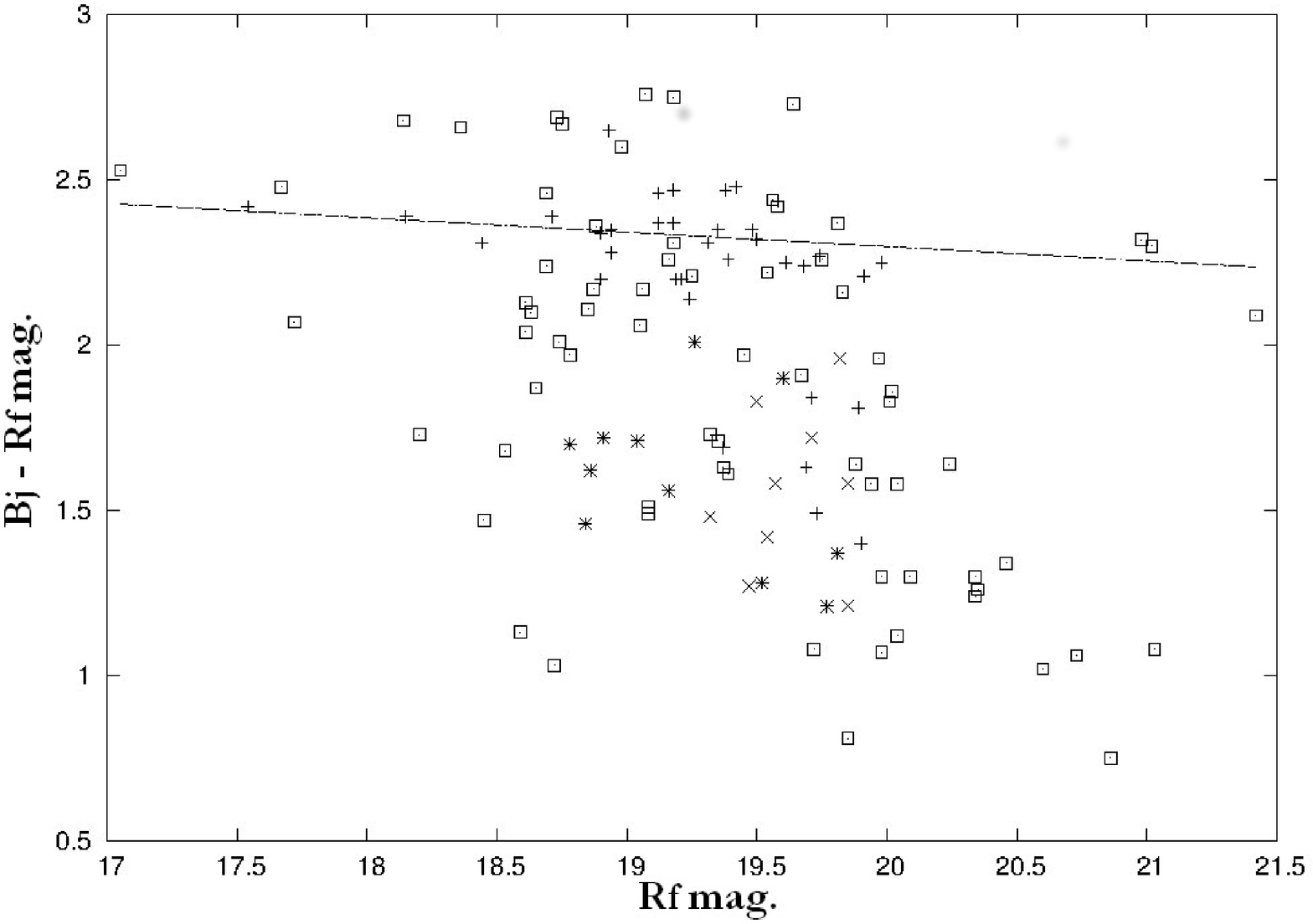}
\par\end{centering}
\caption{the $B_{j}-R_{f}$ diagram as a function of $R_{f}$ for the AC114 cluster member galaxies within an Abell radius of 
5.3 arcmin. (+) and (x) correspond respectively to absorption and emission-lines galaxies observed by Couch \& Sharples 
(1987), while squares and (*) are absorption and emission-lines galaxies from our spectroscopic sample.}
\end{figure*}

\begin{table*}
\begin{minipage}{15 cm}
\caption{Positions and redshifts of foreground, main and distant background structures of AC114.}
\label{tab:S-G-results}
\begin{tabular}{|lcccccc|}
\hline
\hline
Ident & R.A. & Dec   & $\mathrm{z_{cl}}$  & $N_{gal}$ & $N_{gal}$$^($\footnote{~After running the shifting gapper procedure
to remove interlopers.}$^)$ & Probability \\
& & & & 0.5/h~Mpc & 2.5/h~Mpc \\
\hline
AC114 main &  22 58 45.6   & -34 48 00 & 0.317242 & 98 & 177 & 0.99\\ % 171 & 1.00\\
AC114 ~01   &  22 58 43.0   & -34 46 34 & 0.301921  & 4 &   17  & 0.77\\ % 22 58 43.0   & -34 46 34 &  0.301921 &  45 & 0.77\\
AC114 ~02   &  22 59 06.2   & -34 46 16 & 0.409360  & 3 &   10$^($\footnote{~Not confirmed by shifting gapper.}$^)$  & 0.60 \\
\hline
\hline
\end{tabular}
\end{minipage}
\end{table*}

\section{Dynamical mass determinations of AC114}\label{Mass}
The mass of a cluster can be  computed in several ways (Sereno \etal 2010). However, dynamical methods are based on the
asumption that the cluster is close to equilibrium. In the case of AC114, as the foreground structure overlaps partially
the main one, we considered the cluster as a whole. Assuming that galaxies are test particles orbiting 
in a dark matter spherical potential (Binney \& Tremaine 1987), the expression for the virial mass of a cluster is:

\begin{equation}
M_{v}= \frac{3\pi}{2}\frac{\sigma^{2}{\it R}_{Pv}}{G} - {\it C}_{Pr}
\end{equation}

Here ${\it R}_{Pv}$ is the projected virial radius of $N$ observed galaxies and 

\begin{equation}
{\it R}_{Pv}=\frac{N(N-1)}{\Sigma_{i \succ j}R^{-1}_{ij}}
\end{equation}

where $R_{ij}$ is the projected distance between two galaxies given by $i$ and $j$. ${\it C}_{Pr}$ is a surface term 
which accounts that the system is not entirely enclosed in the observational sample with:

\begin{equation}
{\it C}_{Pr} = 4\pi r^{3}_{200}\frac{\rho(r_{200})}{\int^{r_{200}}_{0} 4 \pi r^{2}\rho dr}(\frac{\sigma_{r}(r_{200})}{\sigma(\prec r_{200})})^2
\end{equation}

where $\sigma_{r}(r_{200})$ is the radial velocity dispersion at $r_{200}$ and $\sigma(\prec r_{200})$ is the enclosed total velocity
dispersion within $r_{200}$ (Girardi \etal 1998). On average, for clusters observed within a radius of 1.5/h~Mpc, this
correction is $\simeq 16\%$ (Biviano \etal 2006). The cluster AC114 has a redshift $z= 0.31665$ and a line-of-sight
dispersion $\sigma= 1893\kms$ with $N$=~265 galaxies cluster members between $88500$ and $102000 \kms$. We obtain:
$M_{200}= (4.3 \pm 0.7) \times 10^{15}$M$_{\odot}/h$ enclosed within the virial radius. Our mass estimate is close to 
that derived by Sereno \etal (2010). For a mean redshift $z= 0.3153$ and a velocity dispersion
$\sigma= 1900\kms$ they obtained $M_{200}= (3.4 \pm 0.8) \times 10^{15}M_{\odot}/h$.

The mass estimate can also be computed from the 3D intrinsic velocity dispersion $\sigma_{v}$ within $r_{200}$
(Biviano \etal 2006) with:

\begin{equation}
M_{v} \equiv (1.5 \pm 0.02)(\frac{\sigma_{v}}{10^{3}\kms})^{3} \times 10^{14}~M_{\odot}/h
\end{equation}

The intrinsic velocity dispersion $\sigma_{v}$ is corrected from the velocity dispersion profile following Figure~4
of Biviano \etal (2006) which gives for AC114 $M_{v}= (5.4 \pm 0.7 \pm 0.6) \times 10^{15}$M$_{\odot}/h$. Here the first
error is statistical, and the second one reflects the theoretical uncertainty on the relation (Sereno \etal 2010).
Sereno \etal (2010) use this method to find a similar mass: $M_{v}= (4.8 \pm 0.8 \pm 0.6) \times 10^{15}~M_{\odot}/h$. 
Note that Sereno \etal (2010) derived another mass estimate using the concentration parameter from a triaxial radius
$r_{200}$ such that the mean density contained within an ellipsoid of semi-major axis $r_{200}$ is 200~times the critical
density at the halo redshift. This method yielded a mass estimate $M_{200}= (1.3 \pm 0.9) \times 10^{15}M_{\odot}/h$.

De Filippis \etal (2004) derives X-ray masses from {\it Chandra} images with:

$M_{tot}(1Mpc)= (4.5 \pm 1.1) \times 10^{14} \times h^{-1}_{72}M_{\odot}$

$M_{gas}(1Mpc)= (8.4 \pm 2.6) \times 10^{13} \times h^{-5/2}_{72}M_{\odot}$

They compare their mass estimates with those obtained from weak - and strong lensing by Natarajan \etal (1998)
and conclude that they are in remarkably good agreement. Table~4 summarizes the available mass estimates from 
Natarajan \etal (1998), De Filippis \etal (2004), Sereno \etal (2010) and the present work, in units of 
$10^{14}$M$_{\odot}$ as a function of the radius in Mpc; the method used for mass determination is also indicated.
We note that, despite the fact the mass determinations are made with a wide range of radial extents, they all
appear to be fairly robust measurements with small relative errors, independent of the method used.
If systematic biases exist, they have a weak influence on the results. As quoted in Biviano \etal (2006),
projection effects have a significant impact on the reliability of cluster mass estimates through the inclusion of
interlopers among the samples of presumed cluster members. In the present work we have only considered
galaxies that are securely members of the main cluster and located inside the Abell radius. The presence
of one foreground structure along the line-of-sight may lead to a higher level of uncertainty in the mass estimate
presented.

%\makeatletter\if@referee\renewcommand\baselinestretch{1.0}\fi\makeatother
\begin{table*}[]
\caption{AC114 mass determinations as a function of the radius with different methods.}
\small
\begin{tabular}{ccccl}
\hline
\hline
{\bf Radius (Mpc)} & {\bf mass $10^{14}$M$_{\odot}$/h} & {\bf error bar} & {\bf reference} & {\bf method} \\
\hline
0.075  &  0.42 & 0.01 & Natarajan \etal (1998)    & lensing \\
0.075  &  0.44 & 0.06 & De Philippis \etal (2004) & X-ray \\
0.075  &  0.40 & 0.03 & Sereno \etal (2010)       & ICM, gal. halos, DM \\
0.150  &  1.20 & 0.15 & Natarajan \etal (1998)    & lensing \\
0.150  &  1.12 & 0.17 & De Philippis \etal (2004) & X-ray   \\
0.150  &  1.12 & 0.09 & Sereno \etal (2010)       & ICM, gal. halos, DM \\
0.500  &  4.0  & 0.04 & Natarajan \etal (1998)    & lensing \\
0.500  &  4.7  & 0.10 & De Philippis \etal (2004) & X-ray   \\
1.000  &  4.5  & 0.11 & De Philippis \etal (2004) & $M_{tot}$ \\
3.98   & 34.0  & 8.0  & Sereno \etal (2010)       & Virial \\
3.98   & 43.0  & 7.0  & this work                 & Virial \\ 
3.98   & 48.0  & 8.0  & Sereno \etal (2010)       & Biviano \etal (2006) \\
3.98   & 54.0  & 7.0  & this work                 & Biviano \etal (2006) \\ 
\hline
\end{tabular}
\end{table*}

%\begin{figure*}
%\begin{centering}
%\includegraphics[angle=-90,width=0.9\columnwidth]{massAC114.ps}
%\par\end{centering}
%\caption{AC114 dynamical mass estimates in units of $10^{14}M_{\odot}$ as a function
%of the radius in Mpc.}
%\end{figure*}

\section{Summary and conclusions}\label{Conclusion}

- We have presented a dynamical analysis of the galaxy cluster AC114 based on a catalogue of 524~velocities 
of which 169~(32\%) are newly obtained at the European Southern Observatory with the VLT and the VIMOS spectrograph.

- We obtained an improved mean redshift value $z= 0.31665 \pm 0.0008$ and velocity dispersion 
$\sigma= 1893^{+73}_{-82}\kms$. The cluster has a very elongated main radial filament spanning
$12000 \kms$ in redshift space.  

- { Using Katgert \etal (1996) method, we have been able to detect two secondary structures although at low level
of significance. A radial foreground one is detected within the central 0.5/h~Mpc radius, recognizable as a redshift group
at the same central redshift value. The galaxies taking part of a background structure are sparsely distributed
background to the main AC114, so that no conclusion can be given on a bound structure}.

- AC114 is an archetype Butcher-Oemler galaxy cluster. We analyzed the color distribution for this galaxy cluster 
and identify the sequence of red galaxies which is well separated from the blue galaxies. The latter subset contains
44\% of confirmed members of the cluster, reaching magnitudes as faint as $R_{f}$= 21.1 (1.0 magnitude fainter than
previous studies).

- From the spectroscopic data for $N$=~265 galaxies cluster members we derive a dynamical mass
$M_{200}= (4.3 \pm 0.7) \times 10^{15}$M$_{\odot}$/h for AC114 and $M_{v}= (5.4 \pm 0.7 \pm 0.6) \times 10^{15}$M$_{\odot}$/h
from the intrinsic velocity dispersion out to a radius of 3.98/h~Mpc.

The next phase of this study of AC114 will be to obtain direct metallicities of emission-line galaxies
(Saviane \etal 2014). The redshift of AC114 puts it near the limit where VIMOS can be used to derive oxygen
abundances via the $T_{e}$ method, and this cluster is a well-known gravitational lens, with a number of photometric
data sets already existing. At such a redshift, all important emission lines still fall in the optical range, and the
Universe is $\simeq$ 70\% its current age, so we can expect a factor 1.4 increase in $Z$ since that time,
or 0.14~dex in $[m/H]$. A preliminary presentation of our metallicity analysis can be seen in Saviane \etal (2014). 

\section*{Acknowledgements} We thank ESO-VLT staff for their assistance during the observations (program ID~083.A-0566A), 
and DP thanks ESO in the context of the {\it Visiting Scientists program} for its hospitality at Santiago (Chile). We also thank the referee for their very advised comments.


\begin{thebibliography}{}

\bibitem[Abell \etal (1989)]{Abell89} Abell G.O., Corwin H.G., Olowin R.P.,1989,
ApJS, 70, 1

\bibitem[Adami \etal (1998]{Adami98} Adami C., Mazure A., Biviano A., Katgert P., Rhee G.,1998,
A\&A, 331, 439

\bibitem[Allen (1998)]{Allen98} Allen S.W., 1998, MNRAS, 296, 392

\bibitem[Allen (2000)]{Allen00} Allen S.W., 2000, MNRAS, 315, 269

%\bibitem[Arimoto (1987)]{Arimoto07} Arimoto N., Yoshii Y., 1987, AA 173, 23.

\bibitem[Beers \etal (1990)]{Beers90} Beers T.C., Flynn K., Gebhardt K.,
1990, AJ, 100, 32

\bibitem[Binney (1987)]{Binney87} Binney J., Tremaine S., 1987, Galactic dynamics. Princeton Univ. 
Press, Princeton, NJ 

\bibitem[Biviano \etal (2006)]{Biviano06} Biviano A., Murante G., Borgani S., Diaferio A., Dolag K.,
Girardi M., 2006, A\&A, 456, 23

%\bibitem[Bresolin (2011a)]{Bresolin11a} Bresolin F., 2011, ApJ 730, 129.

%\bibitem[Bresolin (2011b)]{Bresolin11b} Bresolin F., 2011, ApJ 729, 56. 

\bibitem[Butcher (1978)]{Butcher78} Butcher H., Oemler A., 1978, ApJ, 219, 18

\bibitem[Campusano \etal (2001)]{Campusano01} Campusano L.E., Pell\`o R., Kneib J.P.,
Le Borgne J.F., Fort B., Ellis R.S., Mellier Y., Smail Y., 2001, A\&A, 378, 394

\bibitem[Colless \etal (2003)]{Colless03} Colless M., Peterson B.A., Jackson C., Peacock J.A.,
Cole S., Norberg P., Baldry I.K., Baugh C.M., Bland-Hawthorn J., Bridges T., Cannon R., Collins C.,
Couch W., Cross N., Dalton G., De Propris R., Driver S.P., Efstathiou G., Ellis R.S., Frenk C.S.,
Glazebrook K., Lahav O., Lewis I., Lumsden S., Maddox S., Madgwick D., Sutherland W., Taylor K.,
2003, preprint (arXiv:0306581) 

%\bibitem[Couch \etal (1984)]{Couch84} Couch W.J., Newell E.B., 1984, ApJS 56, 143.

\bibitem[Couch \& Sharples (1987)]{Couch87} Couch W.J., Sharples R.M., 1987, MNRAS, 229, 423

\bibitem[Couch \etal (1998)]{Couch98} Couch W.J., Barger A.J., Smail I., Ellis R.S., Sharples R.M.,
1998, ApJ, 497, 518

\bibitem[Couch \etal (2001]{Couch01} Couch W.J., Balogh M.I., Bower R.G., Smail I.,
Glazebrook K., Taylor M., 2001, ApJ, 549, 820

\bibitem[Cypriano \etal (2005)]{Cypriano05} Cypriano E.S., Lima Neto G.B., Sodr\'e Jr. L., 
Kneib J.P., Campusano L.E., 2005, ApJ, 630, 38

\bibitem[Czoske \etal (2001)]{Czoske01} Czoske O., Kneib J.P., Soucail G., Bridges T.J., Mellier Y.,
Cuillandre J.C., 2001, A\&A, 372, 391

\bibitem[Czoske \etal (2002)]{Czoske02} Czoske O., Moore B., Kneib J.P., Soucail G., 2002,
A\&A, 386, 31

%\bibitem[Danese \etal (1980)]{Danese80} Danese L., De Zotti G., di Tullio G., 1980,
%A\&A, 82, 322

\bibitem[De Filippis \etal (2004)]{Defilippis04} De Filippis E., Bautz M.W., Sereno M.,
Garmire G.P., 2004, ApJ, 611, 164

\bibitem[Diaferio (1997)]{Diaferio97} Diaferio A., Geller M.J., 1997, ApJ, 481, 633

\bibitem[Diaferio \etal (2001)]{Diaferio01} Diaferio A., Kauffmann G., Balogh M. L., White S. D. M.,
Schade D.,  Ellingson E., 2001, MNRAS, 323, 999

\bibitem[Dressler Shectman (1988)]{Dressler88} Dressler A., Shectman S.A., 1988, AJ, 95, 284

\bibitem[Fadda \etal (1996)]{Fadda96} Fadda D., Girardi M., Giuricin G., Mardirossian F., Mezzetti M.,
1996, ApJ, 473, 670

\bibitem[Feigelson Babu (2013)]{Feigel13} Feigelson E.D., Babu G.J., 2013, Modern Statistical
Methods for Astronomy, Cambridge University Press

\bibitem[Finn \etal (2004)]{Finn04} Finn R.A., Zaritski D., McCarthy Jr. D.W.: 2004, ApJ, 604, 141

\bibitem[Girardi \etal (1998)]{Girardi98} Girardi M., Giuricin G., Mardirossian F., Mezzetti M., 
Boschin W., 1998, ApJ, 505, 74

\bibitem[Gullieuszik \etal (2009)]{Gullieuszik09} Gullieuszik M., Held E.V., Saviane I., Rizzi L.,
2009, A\&A, 500, 735

\bibitem[Hambly \etal (2001)]{Hambly01} Hambly N.C., Irwin M.J., MacGillivray H.T., 2001,
MNRAS, 326, 1295

\bibitem[Hamuy \etal (1992)]{Hamuy92} Hamuy M., Walker A.R., Suntzeff N.B., Gigoux P., Heathcote S.R.;
Phillips M.M., 1992, PASP, 104, 553

\bibitem[Hamuy \etal (1994)]{Hamuy94} Hamuy M., Suntzeff N.B., Heathcote S.R., Walker A.R.,
Gigoux P., Phillips M.M., 1994, PASP, 106, 566

%\bibitem[Held \etal (2010)]{Held10} Held E.V., Gullieuszik M., Rizzi L., Girardi L.,
%Marigo P., Saviane I., 2010, MNRAS 404, 1475.

%\bibitem[Hong \etal (2010)]{Hong10} Hong S. A., Rosenberg J. L., Ashby M. L. N., Salzer
%J.J., 2010, ApJ 717, 503.

\bibitem[Jones \etal (2009)]{Jones09} Jones D.H., Read M.A., Saunders W., Colless M., Jarrett T.,
Parker Q.A., Fairall A.P., Mauch T., Sadler E.M., Watson F.G., Burton D., Campbell L.A., Cass P., 
Croom S.M., Dawe J., Fiegert K., Frankcombe L., Hartley M., Huchra J., James D., Kirby E., Lahav O.,
Lucey J., Mamon G.A., Moore L., Peterson B.A., Prior S., Proust D., Russell K., Safouris V., 
Wakamatsu K., Westra E., Williams M., 2009, MNRAS, 399, 683

\bibitem[Just \etal (2010)]{Just10} Just D.W., Zaritsky D., Desai V., Rudnick G., 2010,
ApJ, 711, 192

\bibitem[Katgert \etal (1996)]{Katgert96} Katgert P., Mazure A., Perea J., den Hartog R., Moles M.,
Le Fevre O., Dubath P., Focardi P., Rhee G., Jones B., Escalera E., Biviano A., Gerbal D.,
Giuricin G., 1996, A\&A, 310, 8

\bibitem[Kent (1982)]{Kent82} Kent S.M., Gunn J.E., 1982, AJ, 87, 945

\bibitem[Kurtz \etal (1991)]{Kurtz91} Kurtz M.J., Mink D.J., Wyatt W.F., Fabricant D.G., 
Torres G., Kriss G.A., Tonry J.L., 1991, Astron. Soc. Pacific Conf. Ser., 25, 432

\bibitem[Larson (1974)]{Larson74} Larson R.B., 1974, MNRAS, 169, 229

\bibitem[Leccardi (2008)]{Leccardi08} Leccardi A., Molendi S., 2008, A\&A, 486, 359

\bibitem[Leaman \etal (2013)]{Leaman13} Leaman Ryan., Venn K.A., Brooks A.M., Battaglia G., Cole A.,
Ibata R.A., Irwin M.J., McConnachie A.W., Mendel J.T., Starkenburg E., Tolstoy E., 2013, ApJ, 767, 131

\bibitem[Lefevre \etal (1995)]{Lefevre95} Le F\`evre O., Crampton D., Lilly SJ., Hammer F.,
Tresse L., 1995, ApJ, 455, 60 

\bibitem[Lefevre \etal (2003)]{Lefevre03} Le F\`evre O., Saisse M., Mancini D., Brau-Nogue S.,
Caputi O., Castinel L., D'Odorico S., Garilli B., Kissler-Patig M., Lucuix C., Mancini G., Pauget A.,
Sciarretta G., Scodeggio M., Tresse L., Vettolani G., 2003, Proc. SPIE, 4841, 1670  

\bibitem[Lopes \etal (2009]{Lopes09} Lopes P.A.A., de Carvalho R.R., Kohl-Moreira J.L., Jones C.,
2009, MNRAS, 399, 2201

\bibitem[Maddox \etal (1990a)]{Maddox90a} Maddox S.J., Sutherland W.J., Efstathiou G., 
Loveday J., 1990, MNRAS, 243, 692

\bibitem[Maddox \etal (1990b)]{Maddox90b} Maddox S.J., Efstathiou G., Sutherland W.J., 1990,
MNRAS, 246, 433

\bibitem [Mahdavi (2001)]{Mahdavi01} Mahdavi A., Geller M.J., 2001, ApJ, 554, L129

\bibitem[Martini \etal (2006)]{Martini06} Martini P., Kelson D.D., Kim E., Mulchaey J.S., Athey A.A.,
2006, ApJ, 644, 116

\bibitem[Martini \etal (2007)]{Martini07} Martini P., Mulchaey J.S., Kelson D.D., 2007, AJ, 664, 761

\bibitem[Mellier (1999)]{Mellier99} Mellier Y., 1999, ARA\&A, 37, 127

\bibitem[Mink \etal (1995)]{Mink95} Mink D.J., Wyatt W.F., 1995, in Shaw R. A., Payne H. E., Hayes J. J. E., 
eds, ASP Conf. Ser. Vol. 77, Astronomical Data Analysis Software and Systems IV. Astron. Soc. Pac., 
San Francisco, p. 496

%\bibitem[Monaco \etal (2011)]{Monaco11} Monaco L., Saviane I., Correnti M., Bonifacio P.,
%Geisler D., 2011, A\&A 525, 124. 

%\bibitem[Momany \etal (2012)]{Momany12} Momany Y., Saviane I., Smette A., Bayo A., Girardi
%L., Marconi G., Milone A. P., Bressan A., 2012, A\&A 537, 2.

\bibitem[Natarajan \etal (1998)]{Natarajan98} Natarajan P., Kneib J.P., Smail I., Ellis R.S.,
1998, ApJ, 499, 600

%\bibitem[Pence (1976)]{Pence76} Pence W., 1976, ApJ 203, 39.

\bibitem[Pickles (1985)]{Pickles85} Pickles A.J., 1985, ApJS, 59, 33

\bibitem[Pickles (1998)]{Pickles98} Pickles A.J., 1998, PASP, 110, 863

%\bibitem[Proust \etal (2010)]{Proust10} Proust D., Capelato H. V., Lima Neto G. B., 
%Sodr\'e L., 2010, A\&A 515, 57. 

\bibitem[Regos \& Geller (1989)]{Regos89} Regos E., Geller M.J., 1989, AJ, 98, 755

\bibitem[Reisenegger \etal (2000]{Reisenegger00} Reisenegger A., Quintana H., Carrasco E.R.,
Maze J., 2000, AJ, 120, 523
 
\bibitem[Ribeiro \etal (2013)]{Ribeiro13} Ribeiro A.L.B., de Carvalho R.R., Trevisan M.,
Capelato H.V., La Barbera F., Lopes P.A.A., Schilling A.C., 2013, MNRAS, 434, 784 

\bibitem[Rood \etal (1972)]{Rood72} Rood H.J., Page T.L., Kintner E.C., King I.R., 1972, ApJ, 175, 627
 
%\bibitem[Saviane \etal (2012)]{Saviane12} Saviane I., da Costa G. S., Held E. V., Sommariva
%V., Gullieuszik M., Barbuy B., Ortolani S., 2012, A\&A 540, 27. 

\bibitem[Saviane \etal (2014)]{Saviane14} Saviane I., Yegorova I., Proust D., Bresolin F., Ivanov V.,
Held E.V., Salzer J., Rich R.M., 2014, Mem. S. A. I. Vol. 85, p. 417

\bibitem[Sereno \etal (2010)]{Sereno10} Sereno M., Lubini M., Jetzer Ph., 2010, A\&A, 518, 55

%\bibitem[Sharma \etal (2010)]{Sharma10} Sharma S., Borissova J., Kurtev R., Ivanov V.
%D., Geisler D., 2010, AJ, 139, 878

\bibitem[Smail \etal (1995)]{Smail95} Smail I., Couch W.J., Ellis R.S., Sharples R.S., 1995,
ApJ, 440, 501

\bibitem[Tonry \& Davis(1979)]{Tonry79} Tonry J., Davis M., 1979, AJ, 84, 1511

%\bibitem[Wainer \etal (1976)]{Wainer76} Wainer H., Thissen D., 1976, Psychometrika 41, 9

%\bibitem[Yegorova \etal (2011)]{Yegorova11} Yegorova I. A., Pizzella A., Salucci P., 
2011, A\&A, 532, 105

\bibitem[Zuhone \etal (2009]{Zuhone09} ZuHone J.A., Ricker P.M., Lamb D.Q., Karen Yang H.Y., 
2009, ApJ, 699, 1004

\end{thebibliography}
\end{document}